\def\year{2021}\relax

\documentclass[letterpaper]{article} 
\usepackage{aaai20}  
\usepackage{times}  
\usepackage{helvet} 
\usepackage{courier}  
\usepackage[hyphens]{url}  
\usepackage{graphicx} 
\usepackage{graphicx}  
\usepackage{subcaption}
\usepackage{comment}
\usepackage{verbatim}

\usepackage[utf8]{inputenc}
\usepackage[inline]{enumitem}
\usepackage{booktabs}
\usepackage{amssymb}
\usepackage{amsmath}
\usepackage{multirow}
\usepackage{latexsym}
\usepackage{soul}
\usepackage{array}
\usepackage[boxed,linesnumbered]{algorithm2e}
\usepackage{tikz}
\usepackage{xcolor}

\definecolor{con}{HTML}{0070C0}
\definecolor{lab}{HTML}{FF0000}
\definecolor{tvt}{HTML}{FF7D18}
\definecolor{snp}{HTML}{FFFF00}
\definecolor{b60}{HTML}{7030A0}
\definecolor{ase}{HTML}{38C177}
\definecolor{lch}{HTML}{AB7942}
\definecolor{almond}{rgb}{0.94, 0.87, 0.8}
\definecolor{apricot}{rgb}{0.98, 0.81, 0.69}
\definecolor{yellow}{cmyk}{0,0,0.50,0}

\SetCommentSty{mycommfont}
\newcommand{\hashtag}[1]{\textsf{\scriptsize#1}}

\title{Coordinated Behavior on Social Media in 2019 UK General Election}

\author{\Large \textbf{Leonardo Nizzoli,\textsuperscript{\rm 1}\textsuperscript{\rm 2}\thanks{These authors contributed equally.} Serena Tardelli,\textsuperscript{\rm 1}\textsuperscript{\rm 2}\footnotemark[1]\thanks
{Corresponding author. serena.tardelli@iit.cnr.it} Marco Avvenuti,\textsuperscript{\rm 2} Stefano Cresci,\textsuperscript{\rm 1} Maurizio Tesconi\textsuperscript{\rm 1}}\\ 
\textsuperscript{\rm 1}Institute of Informatics and Telematics, National Research Council, Pisa, Italy.\\
\textsuperscript{\rm 2}Department of Information Engineering, University of Pisa, Italy}
\begin{document}

\maketitle

\begin{abstract}
Coordinated online behaviors are an essential part of information and influence operations, as they allow a more effective disinformation's spread. Most studies on coordinated behaviors involved manual investigations, and the few existing computational approaches make bold assumptions or oversimplify the problem to make it tractable.

Here, we propose a new network-based framework for uncovering and studying coordinated behaviors on social media. Our research
extends existing systems and goes beyond limiting binary classifications of coordinated and uncoordinated behaviors. It allows to expose different coordination patterns and to estimate the degree of coordination that characterizes diverse communities. We apply our framework to a dataset collected during the 2019 UK General Election, detecting and characterizing coordinated communities that participated in the electoral debate. Our work conveys both theoretical and practical implications and provides more nuanced and fine-grained results for studying online information manipulation.
\end{abstract}


\section*{Introduction}
\label{sec:intro}
In recent years, \textit{information} or \textit{influence operations} (IOs) have been frequently carried out on social media to mislead and manipulate. With this terminology, borrowed from the military sphere, social media companies have designated the attempts to spread deceptive content on their platforms, orchestrated by state and non-state actors~\cite{alassad2020combining,weedon2017information}. IOs can take different shapes, target various individuals, online crowds, or communities, and have diverse goals~\cite{starbird2019disinfo}. 
Among the strategic tools used by perpetrators are fake news, propaganda, hateful speech, astroturfing, colluding users (e.g., paid trolls), and automation (e.g., social bots). Since the Donald Trump election and the Brexit referendum in 2016, each of these tools has attracted 
extensive scientific attention. The ongoing endeavors have led to a vast body of work on these issues and a plethora of different solutions for solving them. However, despite the efforts, researchers debate the efficacy of the proposed solutions, and IOs still appear to pose a severe threat to our democracies and societies~\cite{barrett2019disinformation}.

Meanwhile, groundbreaking advances in specific areas of computing are causing profound changes to the online information landscape. Advances in artificial intelligence brought to the rise of deepfakes -- synthetic media where the source has been modified via deep learning techniques. Deepfakes allow crafting arbitrary texts that resemble a target person's writing style or audio and video samples where a target person's face and voice could be made to do or say anything.
Unsurprisingly, these powerful techniques have already been used to create fake news~\cite{zellers2019defending} and fake profile pictures for fraudulent accounts\footnote{P. Martineau, ``Facebook Removes Accounts With AI-Generated Profile Photos.'' \emph{Wired}, 20 December 2019. Available at \url{https://www.wired.com/story/facebook-removes-accounts-ai-generated-photos/}}. However, each IO must spread to ``infect'' many users in order to be successful, independently of its aims and tools used to deceive. This goal often mandates extensive and coordinated social media efforts for the campaign to obtain a significant outreach, exert influence, and thus have an impact. In light of this consideration, since 2018, all leading platforms showed great interest in studying \textit{coordinated inauthentic behavior} (CIB)\footnote{N. Gleicher, ``Coordinated Inauthentic Behavior Explained.'' \emph{Facebook}, 6 December 2019. Available at \url{https://about.fb.com/news/2018/12/inside-feed-coordinated-inauthentic-behavior/}}. Despite often appearing together, coordination and inauthenticity are two distinct concepts. For example, activists and other grassroots initiatives typically feature coordinated but authentic behaviors. 
Conversely, a single fake account managed with the intent to mislead might exhibit inauthentic but uncoordinated behavior.
The majority of existing efforts for studying CIB involved plenty of manual investigations, while the computational approaches are still few and far between. Among the challenges, there are the ambiguity and fuzziness of CIB itself: What exactly is a coordinated behavior? What is inauthentic behavior? How many organized accounts are needed for a (meaningful) coordinated behavior to surface? Unfortunately, there are no agreed-upon answers to these questions and, thus, operationalizing such concepts and developing computational methods for their analysis represent open challenges. In particular, so far, no successful attempt has been reported to distinguish between authentic and inauthentic coordination automatically~\cite{vargas2020detection}. Instead, some researchers have recently focused on the more straightforward task
of detecting and studying coordinated behaviors, disregarding intent and authenticity. To this end, the few existing techniques make simple assumptions, such as using fixed thresholds to obtain a binary distinction between coordinated and uncoordinated behaviors~\cite{pacheco2020uncovering}. However, coordination is a complex, non-binary concept, similarly to automation~\cite{cresci2020decade} and inauthenticity~\cite{starbird2019disinformation}.
In the present work, we focus on studying coordinated behavior by combining the advantages and indications of previous preliminary results and moving forward to overcome the drawbacks of a binary approach to coordination.

\subsection{Contributions}

In this work, we go beyond existing approaches for studying coordinated behaviors by proposing a new network-based framework that relaxes previous assumptions, extending and generalizing existing works. Within our framework, we define coordination as a latent, suspicious, or remarkable similarity between any number of users. We do not provide a binary classification of coordinated \textit{vs.} uncoordinated users, but instead, we estimate the extent of coordination.
In summary, with our framework we build a user-similarity network. Then, we obtain the multi-scale backbone of the network by retaining only statistically-relevant links and nodes. Next, we iteratively perform community detection on subsets of increasingly coordinated users. Our approach does not require fixed thresholds for defining coordination. Rather, it allows studying the whole extent of coordination found in the data, from weakly-coordinated to strongly-coordinated users. In particular, the main novelty of our framework is to provide a measure of coordination on a continuous range, instead of separating strongly coordinated accounts and characterizing them apart. Hence, we conclude our analysis by crosschecking our new coordination index with a multifaceted set of network measures, thus characterizing the emerging coordinated communities with unprecedented findings. 
Finally, we test our framework on Twitter data in the context of the 2019 UK General Election (GE) and compare it to previous threshold-based techniques.

Our main contributions are as follows:
\begin{itemize}
    \item We move beyond existing approaches for detecting coordination by proposing a more nuanced, non-binary, network-based framework.
    \item We uncover coordinated communities that operated during the 2019 UK GE, and we discuss them in light of their role in the electoral debate.
    \item We find and discuss different coordination patterns emerging from the behavior of diverse communities. This characterization is made possible by our non-binary approach to coordination, and it demonstrates the power and usefulness of our framework. 
    \item We empirically demonstrate that coordination and automation are orthogonal concepts. Thus, our framework can complement long-studied techniques for detecting automation, manipulation, and inauthenticity.
    \item We perform a comparative evaluation of our framework against existing binary approaches. These results demonstrate the advantages of our nuanced, non-binary framework in terms of its ability to quantify the extent of coordination and to uncover interesting network properties of the coordinated communities.
    \item We create and publicly share a large dataset for the 2019 UK GE, comprising 11M tweets shared by 1.2M users.
\end{itemize}

\section*{Related Work}
\label{sec:relwork}
Due to the many existing challenges, to date, only a few works have attempted to develop computational means to detect and characterize coordinated online behaviors.
Among them, the most similar approach to our present work 
is~\cite{pacheco2020uncovering,pacheco2020unveiling}. Such a work proposed to extract behavioral traces of online activity and use them to build a bipartite network. To do so, they projected the network onto the accounts, obtaining a user similarity network. Then, they filtered low-weight edges, and disconnected nodes, by applying a restrictive and arbitrary similarity threshold. In this way, they deemed the remaining nodes as coordinated and computed the connected components of the filtered network, analyzing each component as a distinct group of coordinated users.
Such a study falls under what we call \textit{threshold-based approaches}, works characterized by a crisp distinction between coordinated and uncoordinated users.
Similarly, in our work, we build a user similarity network. However, the most impactful novelty is that we do not apply a similarity-based filter.
Instead, we iteratively perform community detection at different levels of coordination. In this way, we are able to study the whole extent of coordination among the accounts, uncovering different patterns and dynamics of coordination that would not be visible with previous approaches. Moreover, by measuring coordination on a continuous range, we are able to crosscheck this new coordination index with relevant network properties, obtaining new findings that would not have been possible by characterizing strongly coordinated accounts apart.

Other works experimented with threshold-based approaches.
The work discussed in~\cite{giglietto2020takes,giglietto2020coordinated} focused on a specific instance of CIB. The authors proposed a 2-step process to detect coordinated link-sharing behavior and tested it on a Facebook dataset. 
In the first step, they detected groups of entities that all shared a given link, almost simultaneously. In the second step, the coordinated networks were identified by connecting only those entities that repeatedly shared the same links. Then, they manually assessed inauthenticity by analyzing shared domains and stories. The proposed algorithm required two parameters: one for defining near-simultaneous link sharing, and the other for defining repetitive link sharing. Such parameters represented fixed similarity thresholds used for filtering, bringing along all previous implications and limitations. Assenmacher \textit{et al.} also proposed a 2-step framework for detecting IOs~\cite{assenmacher2020two,assenmacher2020semi}. Initially, they leveraged unsupervised stream clustering and trend detection techniques to social media streams of text, identifying groups of similar users. Then, they proposed applying standard offline analyses, including manual inspection via visualizations and dashboards, for assessing inauthenticity. ~\cite{keller2020political} leveraged a ground-truth of coordinated accounts involved in a disinformation campaign to identify network measures for detecting IOs. The authors concluded that the traces left by coordination among astroturfing agents are more informative than the typical individual account characteristics used for other related tasks (e.g., social bot detection). In addition, they developed an astroturfing detection methodology based on the previously identified coordination patterns. In~\cite{fazil2020socialbots}, the authors proposed a multi-attributed graph-based approach for detecting CIB on Twitter. The authors modeled each user with a 6-dimensional feature vector, computed pairwise similarities to obtain a user similarity graph, and finally applied Markov clustering, labeling the resulting clusters as inauthentic coordinated groups. In~\cite{fazil2020socialbots}, high coordination automatically implied inauthenticity.


Instead of proposing a new technique, the study in~\cite{vargas2020detection} focused on determining the usefulness and reliability of previously-proposed network-based metrics of coordination. Several authors, including some of those previously mentioned, reported positive results for detecting inauthentic behavior via the analysis of suspicious coordination~\cite{ratkiewicz2011detecting,keller2020political,fazil2020socialbots}. 
%
However, the results of~\cite{vargas2020detection} showed that, when evaluated in non-trivial real-world scenarios, such previously proposed approaches could not distinguish between authentic (e.g., activists, fandoms) and inauthentic coordination. These results confirm that coordination and inauthenticity are different concepts, and that high coordination does not necessarily imply inauthenticity.

\section*{Dataset}
\label{sec:dataset}
By leveraging Twitter Streaming APIs, we collected a large dataset of tweets related to the 2019 UK GE. Our data collection covered one month before the election day, from 12 November to 12 December 2019. During that period, we collected all the tweets mentioning at least one hashtag from a list containing the most popular hashtags, some used by the two main parties while others more neutral. Table~\ref{tab:dataset-hashtags} lists all the hashtags used during this phase, their corresponding political leaning (N: Neutral, L: Labour, C: Conservative), and the related data collected. The \textit{tweets} column only counts quoted retweets if the quote text contains one of the hashtags in the table. The remaining quoted retweets are still included in our dataset, but they are not counted in Table~\ref{tab:dataset-hashtags}. 
In addition to the aforementioned hashtags-based collection, we collected all tweets published by the two parties' official accounts and their leaders, together with all the interactions (i.e., retweets and replies) they received. 
Table~\ref{tab:dataset-accounts} shows the accounts and the collected data. Our final dataset for this study is the combination of the data shown in Tables~\ref{tab:dataset-hashtags},~\ref{tab:dataset-accounts}, and quoted retweets, and includes a total of 11,264,820 tweets published by 1,179,659 distinct users. This dataset is publicly available for research purposes\footnote{\url{https://doi.org/10.5281/zenodo.4647893}}.

\begin{table}[t]
	\scriptsize
	\centering
	\begin{tabular}{lcrrr}
		\toprule
		\textbf{hashtag} & \textit{leaning} & \textit{users} & \textit{tweets} \\
		\midrule
		\textsf{\#GE2019}	                & N & 436,356			& 2,640,966			\\
		\textsf{\#GeneralElection19}	    & N & 104,616			& 274,095			\\
		\textsf{\#GeneralElection2019}	    & N & 240,712			& 783,805			\\
		\textsf{\#VoteLabour}	            & L & 201,774			& 917,936			\\
		\textsf{\#VoteLabour2019}	        & L & 55,703			& 265,899			\\
		\textsf{\#ForTheMany}	            & L & 17,859			& 35,621			\\
		\textsf{\#ForTheManyNotTheFew}	    & L & 22,966			& 40,116			\\
		\textsf{\#ChangeIsComing}	        & L & 8,170			& 13,381            \\
		\textsf{\#RealChange}	            & L & 78,285			& 274254			\\
		\textsf{\#VoteConservative}	        & C & 52,642			& 238,647			\\
		\textsf{\#VoteConservative2019}	    & C & 13,513			& 34,195			\\
		\textsf{\#BackBoris}	            & C & 36,725			& 157,434			\\
		\textsf{\#GetBrexitDone}	        & C & 46,429			& 168,911			\\
		\midrule
		\textbf{total}	                    & --& 668,312			& 4,983,499		        \\
		\bottomrule
	\end{tabular}
	\caption{Statistics about data collected via hashtags.}
	\label{tab:dataset-hashtags}
\end{table}

\begin{table}[t]
	\scriptsize
	\centering
	\begin{tabular}{lcrrrr}
		\toprule
		& \multicolumn{2}{c}{\textbf{production}} && \multicolumn{2}{c}{\textbf{interactions}}\\
		\cmidrule{2-3} \cmidrule{5-6}
		\textbf{account} & \textit{leaning} & \textit{tweets} && \textit{retweets} & \textit{replies}\\
		\midrule
		\textsf{@jeremycorbyn}  & L & 788   && 1,759,823    & 414,158   \\
		\textsf{@UKLabour}      & L & 1,002 && 325,219      & 79,932    \\
		\textsf{@BorisJohnson}  & C & 454   && 284,544      & 382,237   \\
		\textsf{@Conservatives} & C & 1,398 && 151,913      & 169,736   \\
		\midrule
		\textbf{total}          & --& 3,642	&& 2,521,499    & 1,046,063 \\
		\bottomrule
	\end{tabular}
	\caption{Statistics about data collected from accounts.}
	\label{tab:dataset-accounts}
\end{table}

\begin{figure*}[t]
    \centering
    \includegraphics[width=1\textwidth]{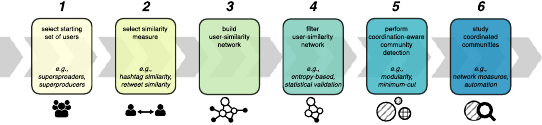}
    \caption{Overview of the proposed framework for studying coordinated behavior.}
    \label{fig:method}
\end{figure*}

\section*{Method overview}
\label{sec:method}
In this section, we describe our network-based framework for detecting coordinated behaviors. Our detailed methodology is composed of the following six main steps, summarized in Figure~\ref{fig:method}:

\begin{enumerate}
    \item \label{itm:step1}\textbf{Select starting set of users.} The first step concerns the selection of those users to investigate. For instance, given a large dataset, one might want to investigate most-active users, such as \textit{superproducers} or \textit{superspreaders}\footnote{\textit{Superproducers} are those users responsible for creating the majority of original content, whereas \textit{superspreaders} are those mostly contributing to spread existing content.}, or all users who tweeted with a particular hashtag, or even all followers of a given prominent user. Whatever the selection criterion, this step returns a list of users to analyze. 
    \item \label{itm:step2}\textbf{Select similarity measure.} 
    As in the previous literature, we leverage unexpected similarities between users as a proxy for coordination. Similarity can be computed in many different ways. Hence, this step deals with the selection of a similarity measure. Examples of valid options are the cosine similarity between user feature vectors encoding account profile characteristics, as done in~\cite{fazil2020socialbots}, or the Jaccard similarity between the sets of hashtags used by each user or between the sets of followings/retweeted accounts.
    \item \label{itm:step3}\textbf{Build user similarity network.} In this step, we compute pairwise user similarities between all users identified at step~\ref{itm:step1}, employing the metric selected at step~\ref{itm:step2}. We leverage user similarities to build a weighted undirected user similarity network $G(E,V,W)$ that encodes behavioral and interaction patterns between users.
    \item \label{itm:step4}\textbf{Filter user similarity network.} When studying real-world datasets of large IOs, the network resulting from step~\ref{itm:step3} can be too big to analyze and even to visualize. Hence, a filtering step is needed. Contrarily to previous works focused on threshold-based approaches, we avoid simple filtering strategies based on fixed edge weight thresholds. We recall that edge weights encode similarity and, to a certain extent, coordination between users. As such, applying a weight threshold $t$ and discarding all edges $e \in E$ whose weight is lower than the threshold ($w(e) < t$) would mean to arbitrarily perform a binary distinction between coordinated behaviors ($w(e) \ge t$) and uncoordinated ones ($w(e) < t$), which is a limiting and theoretically-unmotivated choice. Instead, we propose to use complex network-based multiscale filtering methods, such as any of those discussed in~\cite{garlaschelli2008maximum,serrano2009extracting,tumminello2011statistically}. These techniques retain statistically-meaningful network structures independently on their scale (i.e., edge weight). As such, the network filtering step is not biased towards certain levels of similarity and coordination
    but instead erases network structures that convey limited information, allowing us to focus on meaningful similarities.
    \item \label{itm:step5}\textbf{Perform coordination-aware community detection.} The detection of coordinated groups of users is often achieved via clustering and community detection. 
    Threshold-based approaches consider the filtered user similarity network only to contain highly-coordinated users. Thus, a single run of a community detection algorithm is enough for them to highlight coordinated groups.
    Instead, in our case, the filtered user similarity network still features diverse levels of coordination and needs a more nuanced approach for surfacing coordinated behaviors. As such, we base our approach on an iterative process that examines increasing coordination levels, as shown in Algorithm~\ref{alg:comdet}. We begin by performing community detection on the filtered network resulting from step~\ref{itm:step4}, identifying the set $C_0$ of communities. Then, we apply an increasingly restrictive similarity threshold $t_i$ to edge weights at each iteration, thus removing certain edges and disconnected nodes. We repeat community detection on the obtained subnetwork $G^{e,v}_i$. At each iteration, the community detection algorithm is initialized with the set of communities $C_{i-1}$ found at the previous iteration. This process guarantees that the starting communities are kept, to a certain extent\footnote{Communities may still break or merge, which we account for in our process.}, throughout all the process. As a result of the ``moving'' threshold, we are able to study how the structure and the properties of coordinated communities change across the whole spectrum of coordination. Moreover, the moving threshold implicitly defines a measure for the extent of coordination observed at each iteration, that is for each obtained subnetwork.
    \item \label{itm:step6}\textbf{Study coordinated communities.} To study the structure of coordinated communities and their patterns of coordination, we employ several network measures. In addition, we put communities into context, and we characterize their content production by applying natural language processing techniques. By leveraging our novel approach to the detection of coordinated communities described in step~\ref{itm:step5}, we can obtain these analyses' results as a function of the measured extent of coordination between users.
\end{enumerate}

\IncMargin{1em}
\begin{algorithm}
    \small
    \DontPrintSemicolon
    \SetKwInOut{Input}{input}
    \SetKwInOut{Output}{output}
    \Indm
    \KwData{$G(E,V,W)$ \tcp*{filtered user similarity network}}
    \KwResult{$C$}
    \Indp
    \BlankLine
    \tcc{initialization}
    $C_0 =$ perform community detection on $G$\;
    $C = \langle C_0 \rangle$\;
    $t_0 = \min(w \in W)$\;
    \tcc{detect communities as a function of coordination}
    $i = 1$\;
    \While{$t_{i-1} + \delta_w \le \max(w \in W)$}{
        $t_i = t_{i-1} + \delta_w$ \tcp*{increment threshold by step $\delta_w$}
        $E^- = \{e \in E \; | \; w(e) < t_i\}$ \tcp*{filter out edges}
        $G^e_i = G - E^-$\;
        $V^- = \{v \in V^e_i \; | \; d(v) = 0\}$ \tcp*{filter out nodes}
        $G^{e,v}_i = G^e_i - V^-$ \tcp*{obtain subnetwork $G^{e,v}_i$}
        initialize community detection with $C_{i-1}$\;
        $C_i =$ perform community detection on $G^{e,v}_i$\;
        append $C_i$ to $C$ \tcp*{trace evolving communities}
        $i = i + 1$\;
    }
    \Return $C$\;
    \caption{Coordination-aware community detection. We perform the community detection iteratively at increasing coordination levels, exposing how communities' structure and properties change when imposing increasingly restrictive similarity thresholds.\label{alg:comdet}}
\end{algorithm}
\DecMargin{1em}

In summary, the first three steps of our framework lead to the user similarity network, and they are also present in a similar way in previous threshold-based approaches. In turn, step ~\mbox{\ref{itm:step4}} is where we depart from previous works since we filter the network to retain the most significant, multi-scale structures instead of performing a single-scale analysis on the most similar accounts. 
However, our main novelty is the coordination-aware community detection algorithm of step~\ref{itm:step5}, which allows us to measure coordination at all its extent and study it as a function of other relevant network properties (step~\ref{itm:step6}), which would not have been possible with threshold-based approaches. In the following, we use our framework to uncover and characterize possible coordinated behaviors affecting the 2019 UK General Election. Then, we demonstrate the benefits of our new framework by comparing its findings with those obtained by applying a threshold-based approach.
\section*{Surfacing coordination in 2019 UK GE}
\label{sec:detection}
In the following, we describe how we implemented and applied the aforementioned framework to uncover coordinated behaviors on Twitter related to the 2019 UK GE. This section's content roughly corresponds to steps~\ref{itm:step1} to~\ref{itm:step5} of our methodology, while the next section describes step~\ref{itm:step6} (i.e., analysis of coordinated communities).

\begin{figure}[t]
    \centering
    \includegraphics[trim={4cm 0 0 12cm},width=1\columnwidth]{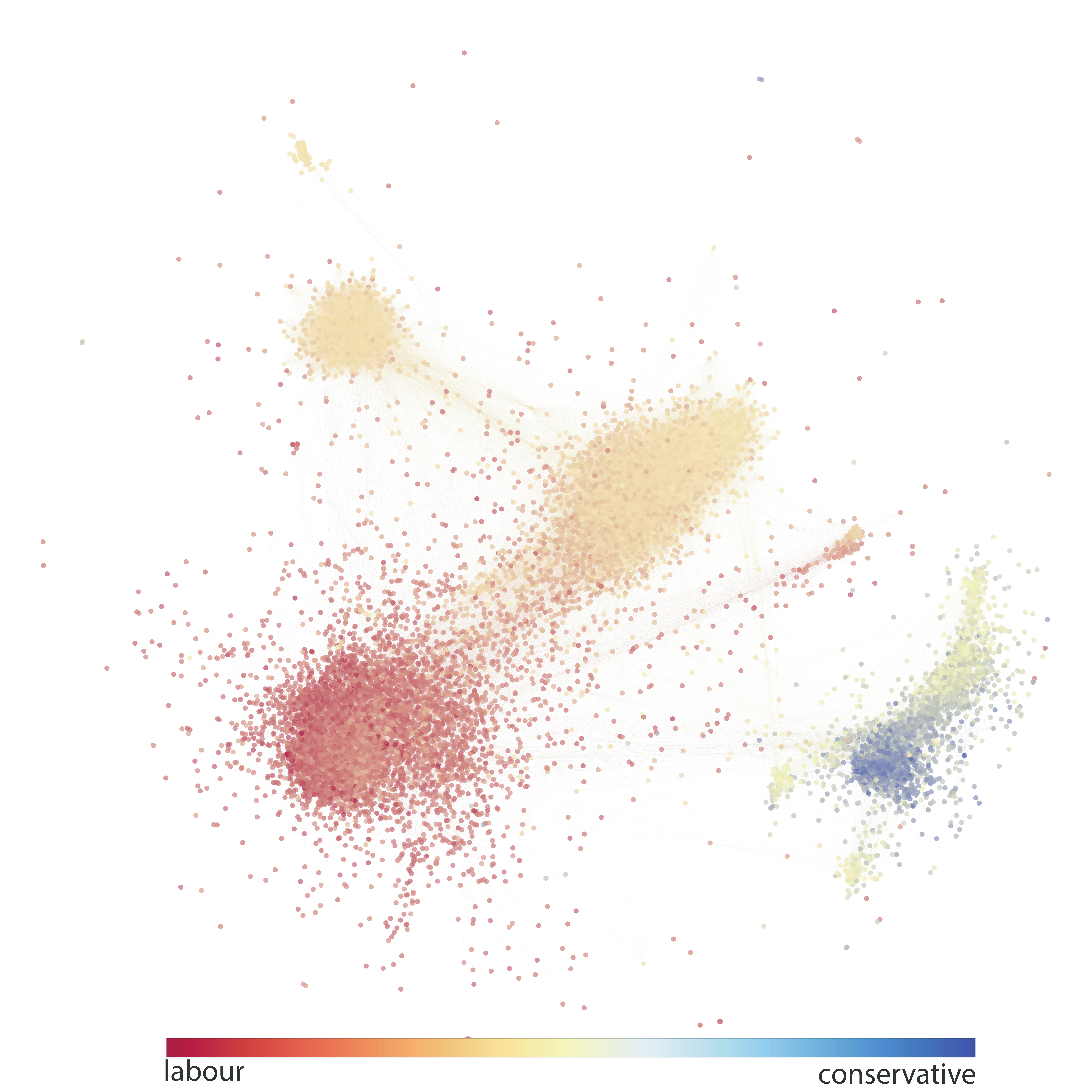}
    \caption{
    The filtered user similarity network of the 2019 UK GE. Nodes represent users involved in the online debate, while edges weight according to the similarity strength between one another, as defined in Step 2. Users are colored according to their political leaning, inferred from the polarity of the hashtags they used, computed using label propagation.
    }
    \label{fig:network}
\end{figure}

\begin{figure}[t]
    \centering
    \includegraphics[width=0.6\columnwidth]{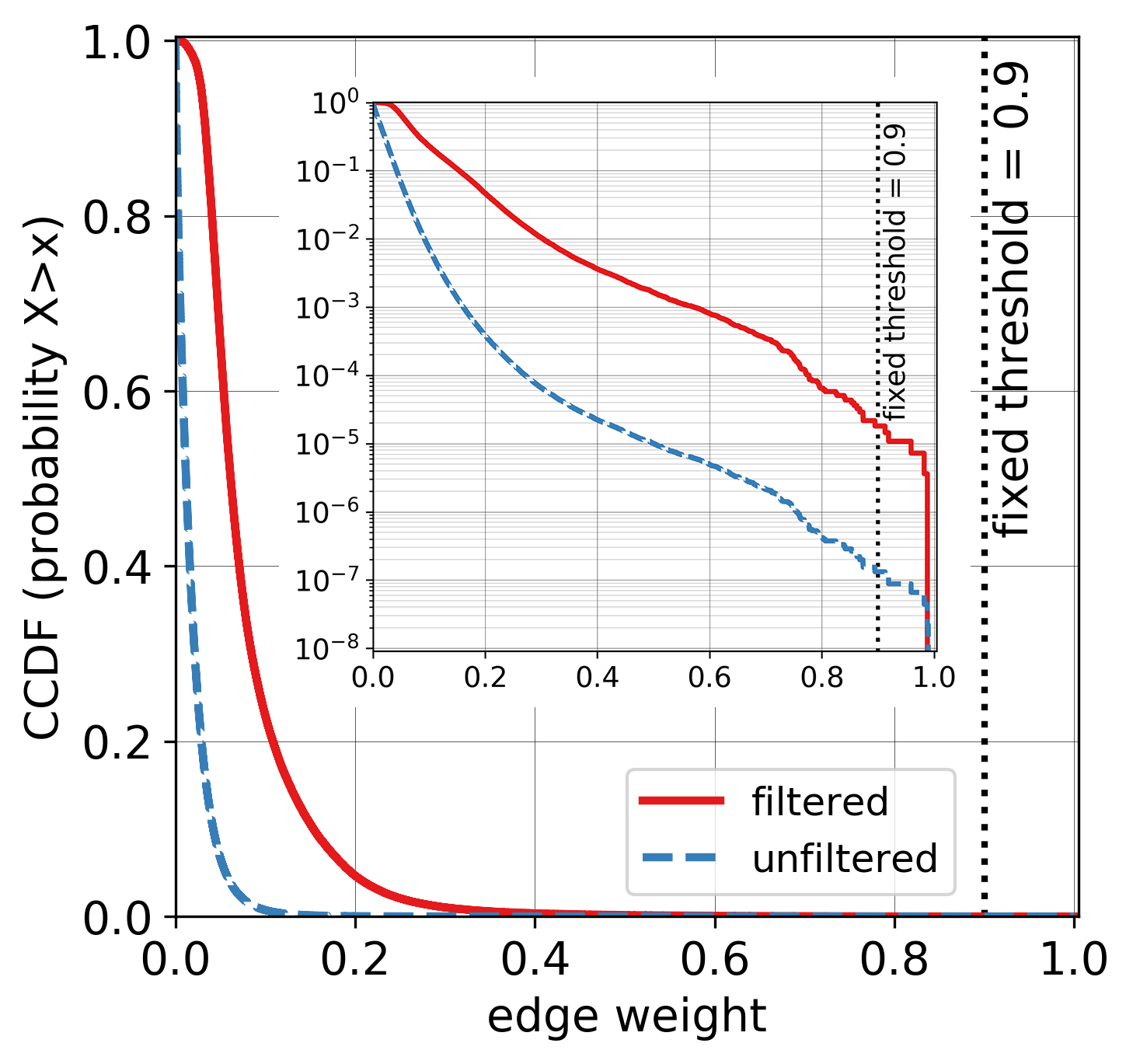}
    \caption{Edge weight distribution of the unfiltered (blue-colored) and filtered (red-colored) user similarity networks. The network filtered using a multiscale filtering method allows us to retain statistically-meaningful network structures independently on their scale (i.e., edge weight), as opposed to cutting at a predetermined, fixed threshold (dotted-black line), which would exclude almost the entire network from the subsequent analysis.}
    \label{fig:edge-weights}
\end{figure}

\textbf{User similarity network.} For our analysis, we focused on the activity of \textit{superspreaders} -- coarsely defined as the most influential spreaders of information, including mis- and disinformation, in online social media~\cite{pei2014searching}. Here, we defined superspreaders as the top 1\% of users that shared more retweets. This definition resulted in the selection of 10,782 users for our analysis. 
Despite representing only the 1\% of all users in the online electoral debate, superspreaders shared the 39\% of all tweets and the 44.2\% of retweets.
Thus, by focusing on them, we investigated the most prolific users and a considerable share of all messages. Next, we characterized each superspreader with a TF-IDF weighted vector of its retweeted tweet IDs. In other words, we modeled the users according to the tweets they retweeted. The TF-IDF weight allows us to reduce the relevance of highly popular tweets in our dataset and emphasize similarities due to retweets of unpopular tweets -- a much more suspicious behavior.~\cite{mazza2019rtbust}. Then, we computed user similarities as the cosine similarity of user vectors. Before studying the network, we applied the technique proposed in~\cite{serrano2009extracting} to retain only statistically-relevant edges, thus obtaining the multiscale backbone of our network, which we exploited for the remaining analyses. The resulting filtered user similarity network contains 276,775 edges and is shown in Figure~\ref{fig:network}. In addition, Figure~\ref{fig:edge-weights} shows the distribution of edge weights in the filtered network. The filtering step preserved the rich, multiscale nature of the network, as opposed to filtering based on a fixed threshold.

\textbf{Political leaning.} In Figure~\ref{fig:network}, nodes are colored based on their political leaning, as inferred from the hashtags that they used. 
In particular, we employed a label propagation algorithm for assigning a polarity score to each hashtag in our dataset. In detail, it is a modified version of the so-called \textit{valence score}, a simple, standard, well-known method from literature adopted in many studies before~\cite{wang2011topic}.
The score for a given hashtag is inferred from its co-occurrences with seeds of known polarity. We used the 13 hashtags in Table~\ref{tab:dataset-hashtags} as the seeds for the label propagation.  Finally, a user's polarity is computed as the term-frequency weighted average of the polarities of the hashtags used by that user. 


\begin{figure*}[t]
    \centering
	\begin{minipage}[b]{1\textwidth}%
        \centering
        \includegraphics[trim={11cm 5cm 4cm 6cm},width=0.6\textwidth]{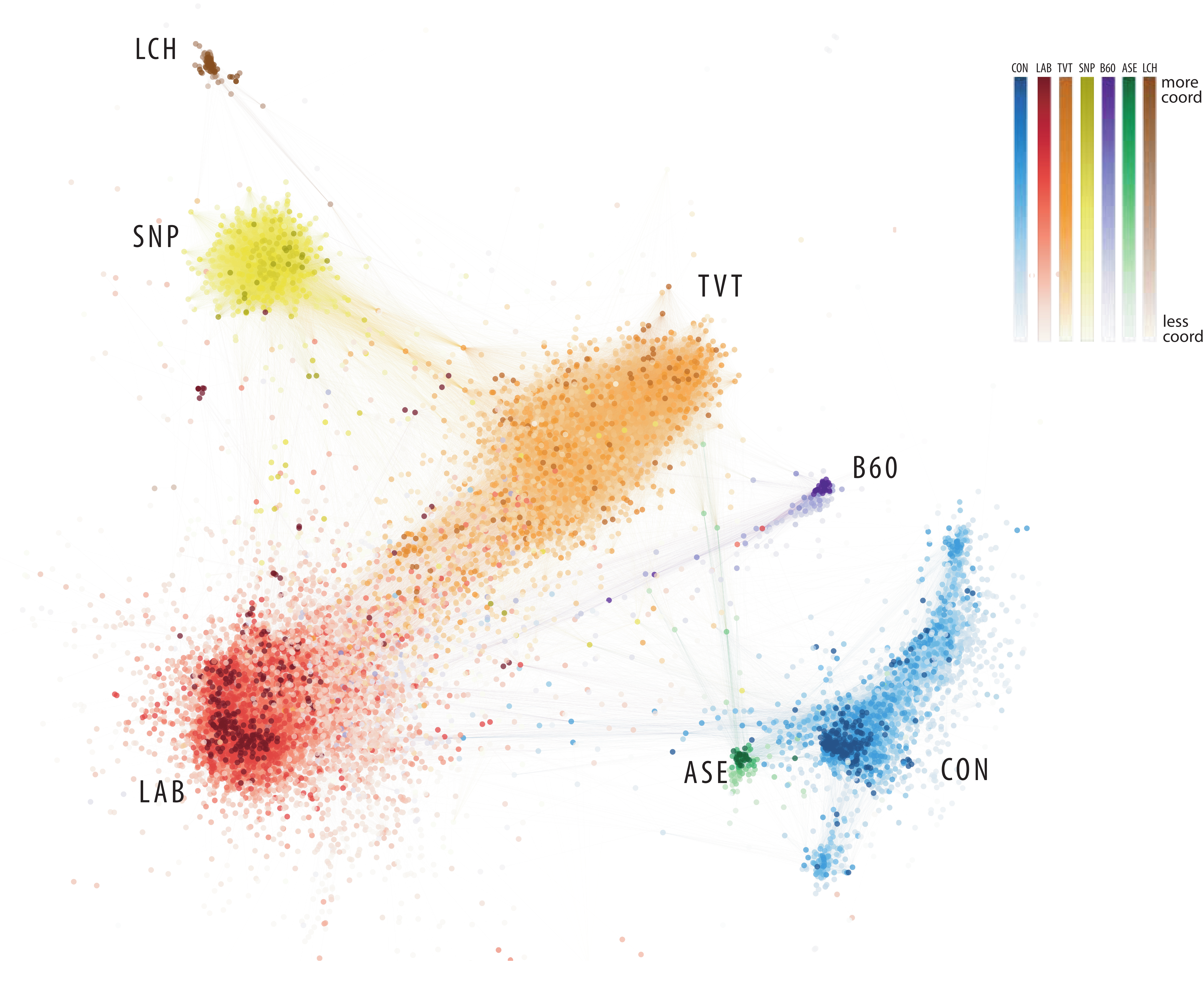}
        
        \caption{Coordinated communities found within the filtered user similarity network. Communities are color-coded. For each color, intensity encodes the extent of coordination. In this way, users exhibiting higher coordination with other users are identified with darker shades of colors.}
        \label{fig:network-communities}
	\end{minipage}\\%
	\medskip
	\begin{minipage}[b]{1\textwidth}%
        \centering
        \begin{subfigure}{.125\textwidth}
            \includegraphics[width=\textwidth]{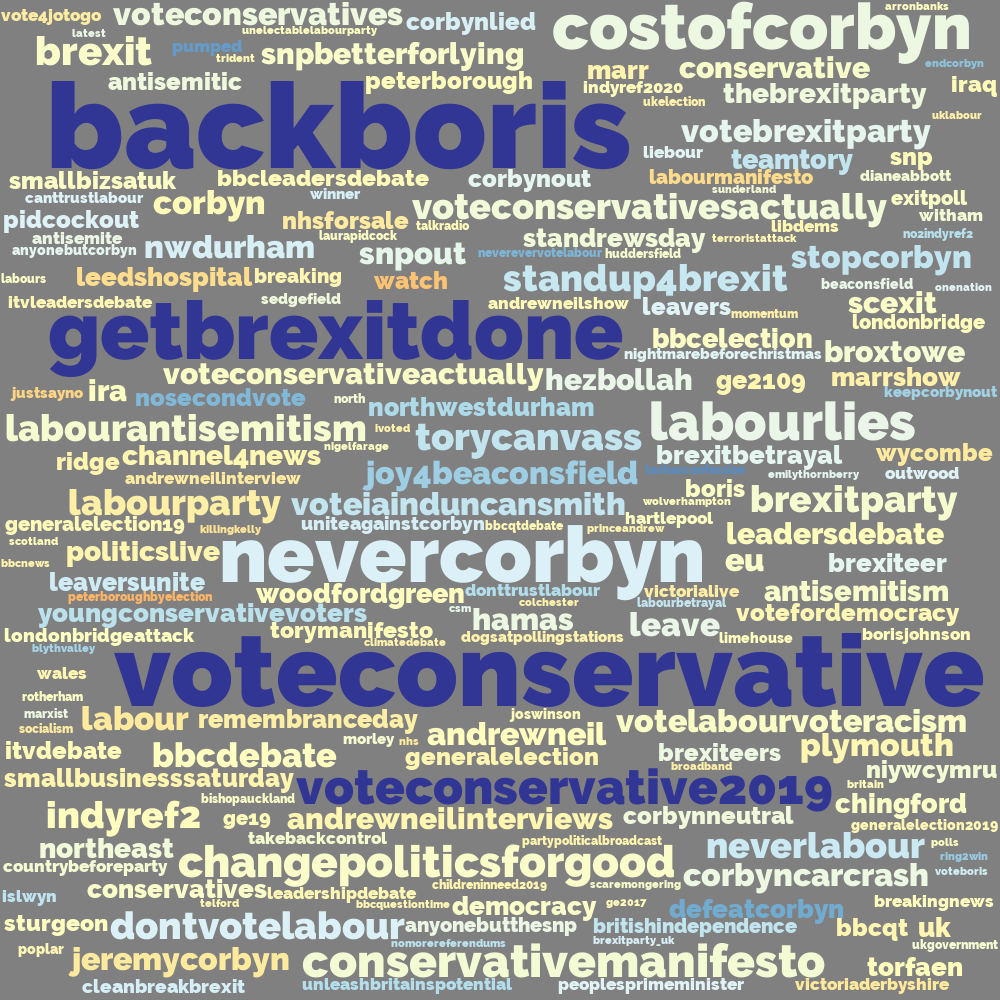}
            \caption{\tikz\draw[black, fill=con] (0,0) circle (.75ex);~\texttt{CON}.\label{fig:network-hashtags-CON}}
        \end{subfigure}\hspace{.015\textwidth}%
        \begin{subfigure}{.125\textwidth}
            \includegraphics[width=\textwidth]{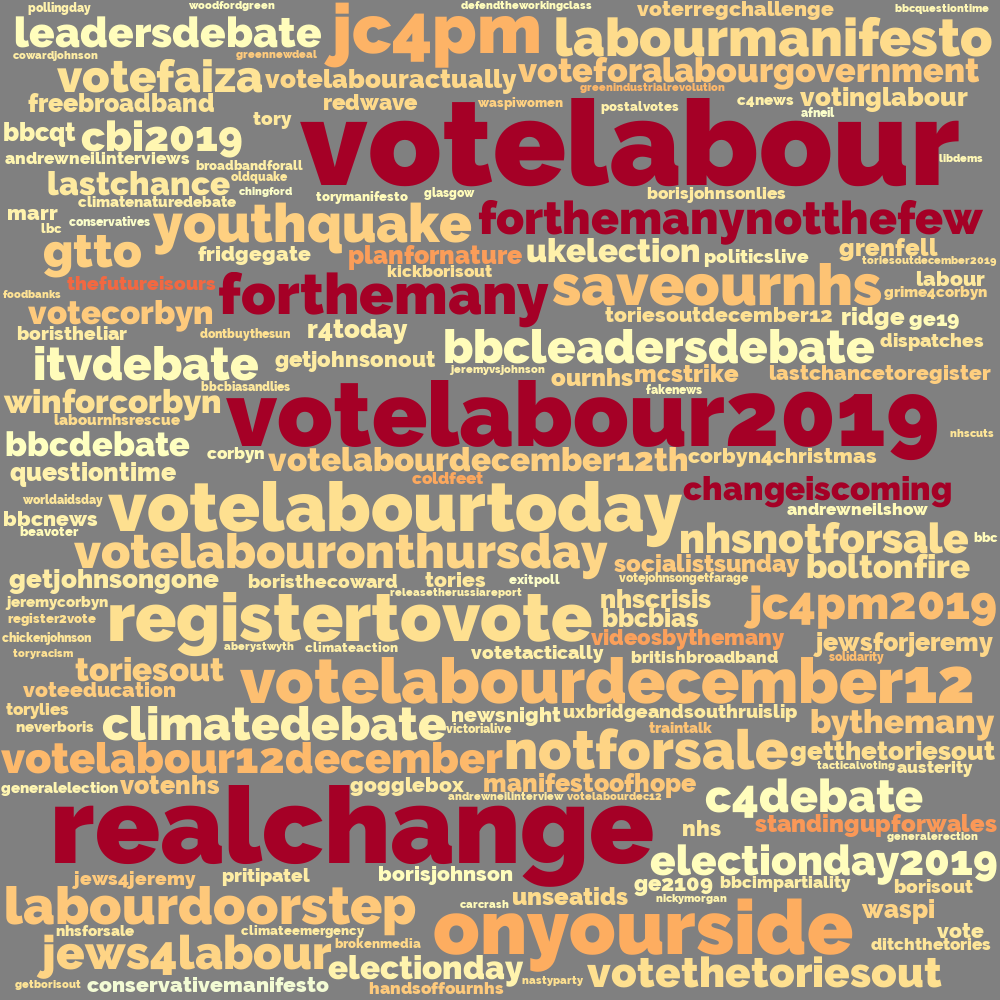}
            \caption{\tikz\draw[black, fill=lab] (0,0) circle (.75ex);~\texttt{LAB}.\label{fig:network-hashtags-LAB}}
        \end{subfigure}\hspace{.015\textwidth}%
        \begin{subfigure}{.125\textwidth}
            \includegraphics[width=\textwidth]{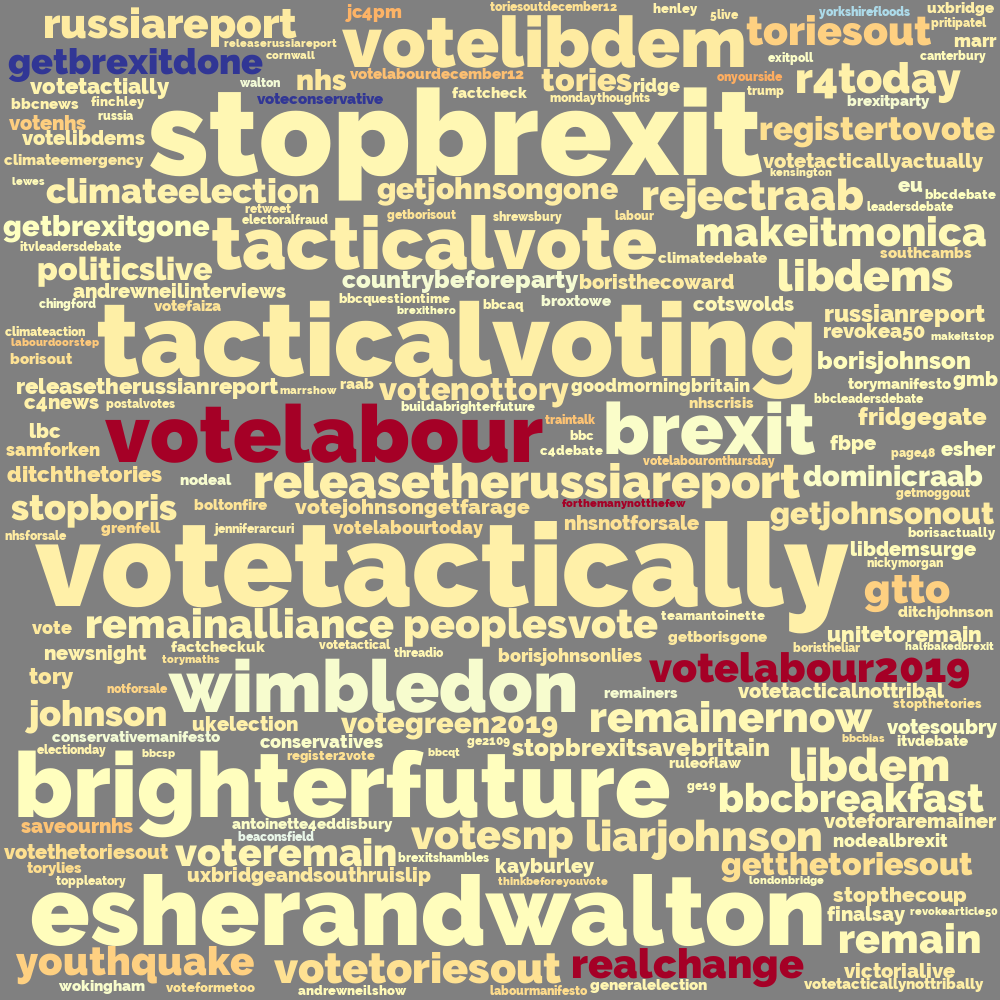}
            \caption{\tikz\draw[black, fill=tvt] (0,0) circle (.75ex);~\texttt{TVT}.\label{fig:network-hashtags-TVT}}
        \end{subfigure}\hspace{.015\textwidth}%
        \begin{subfigure}{.125\textwidth}
            \includegraphics[width=\textwidth]{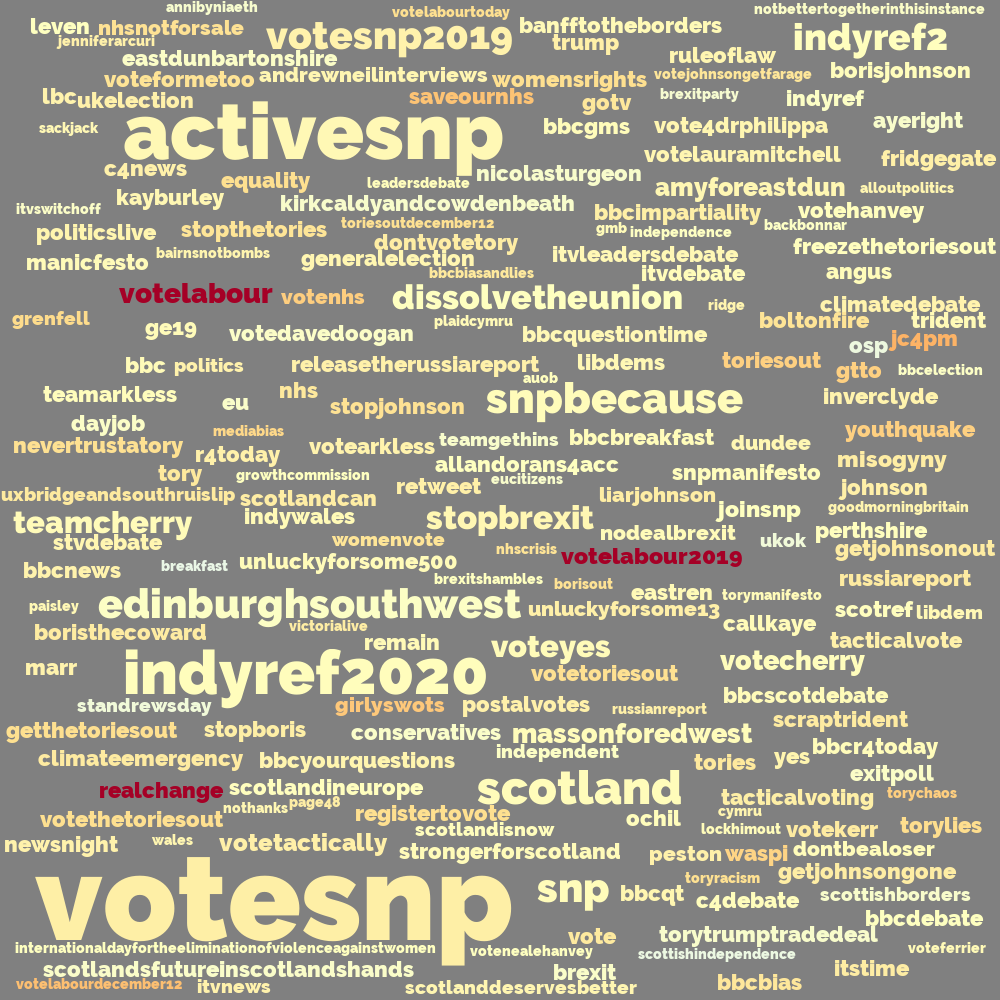}
            \caption{\tikz\draw[black, fill=snp] (0,0) circle (.75ex);~\texttt{SNP}.\label{fig:network-hashtags-SNP}}
        \end{subfigure}\hspace{.015\textwidth}%
        \begin{subfigure}{.125\textwidth}
            \includegraphics[width=\textwidth]{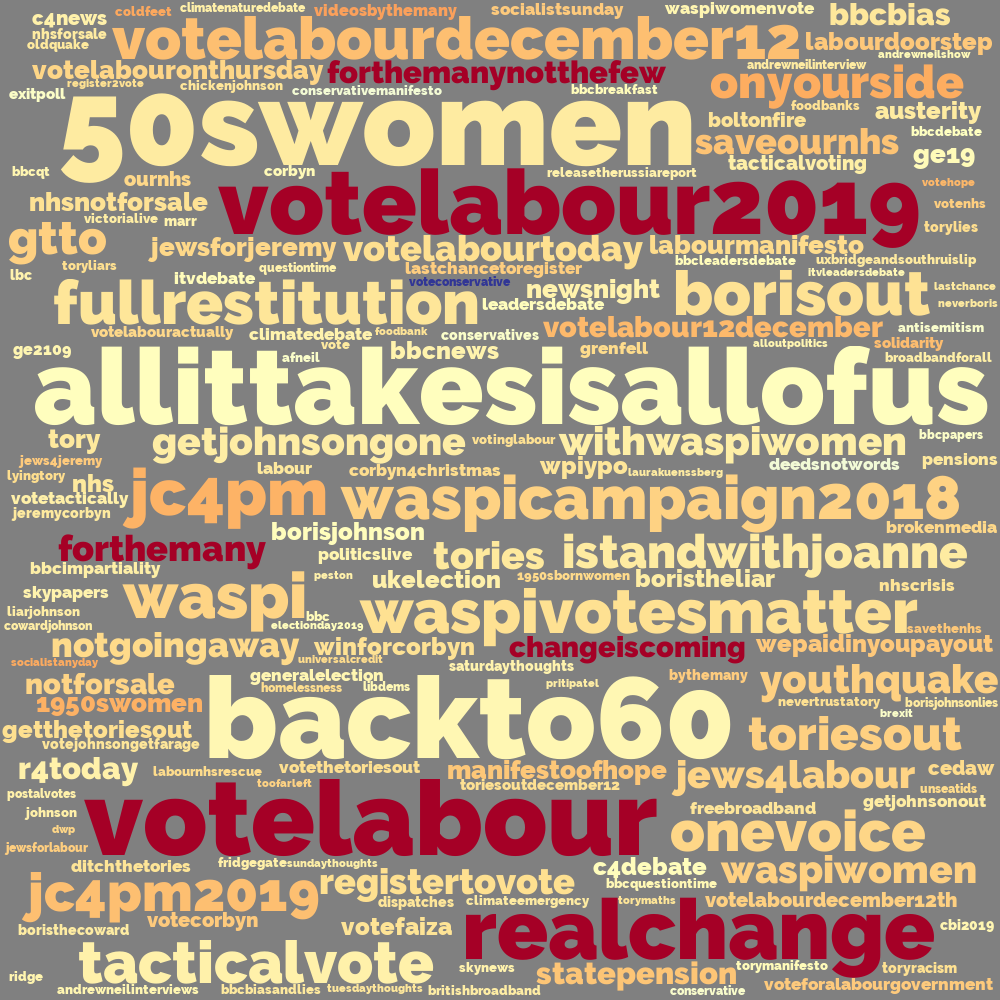}
            \caption{\tikz\draw[black, fill=b60] (0,0) circle (.75ex);~\texttt{B60}.\label{fig:network-hashtags-B60}}
        \end{subfigure}\hspace{.015\textwidth}%
        \begin{subfigure}{.125\textwidth}
            \includegraphics[width=\textwidth]{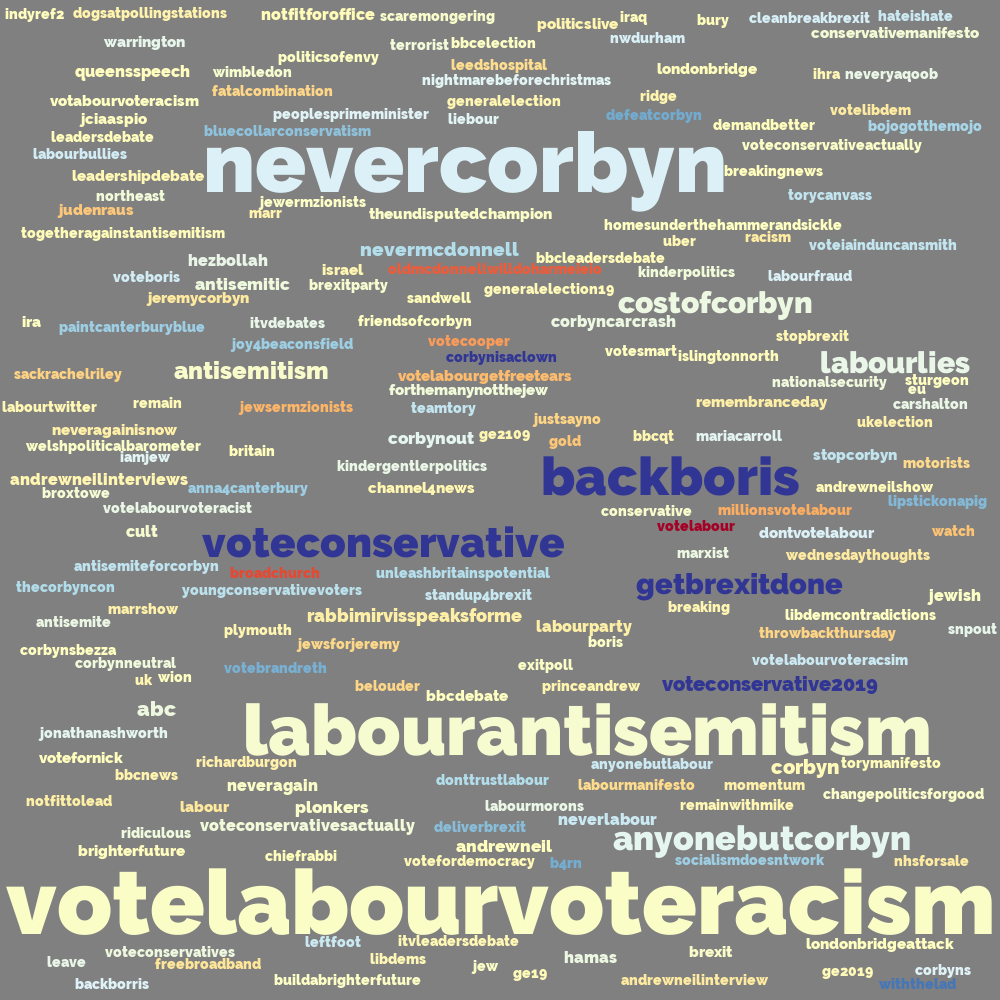}
            \caption{\tikz\draw[black, fill=ase] (0,0) circle (.75ex);~\texttt{ASE}.\label{fig:network-hashtags-ASE}}
        \end{subfigure}\hspace{.015\textwidth}%
        \begin{subfigure}{.125\textwidth}
            \includegraphics[width=\textwidth]{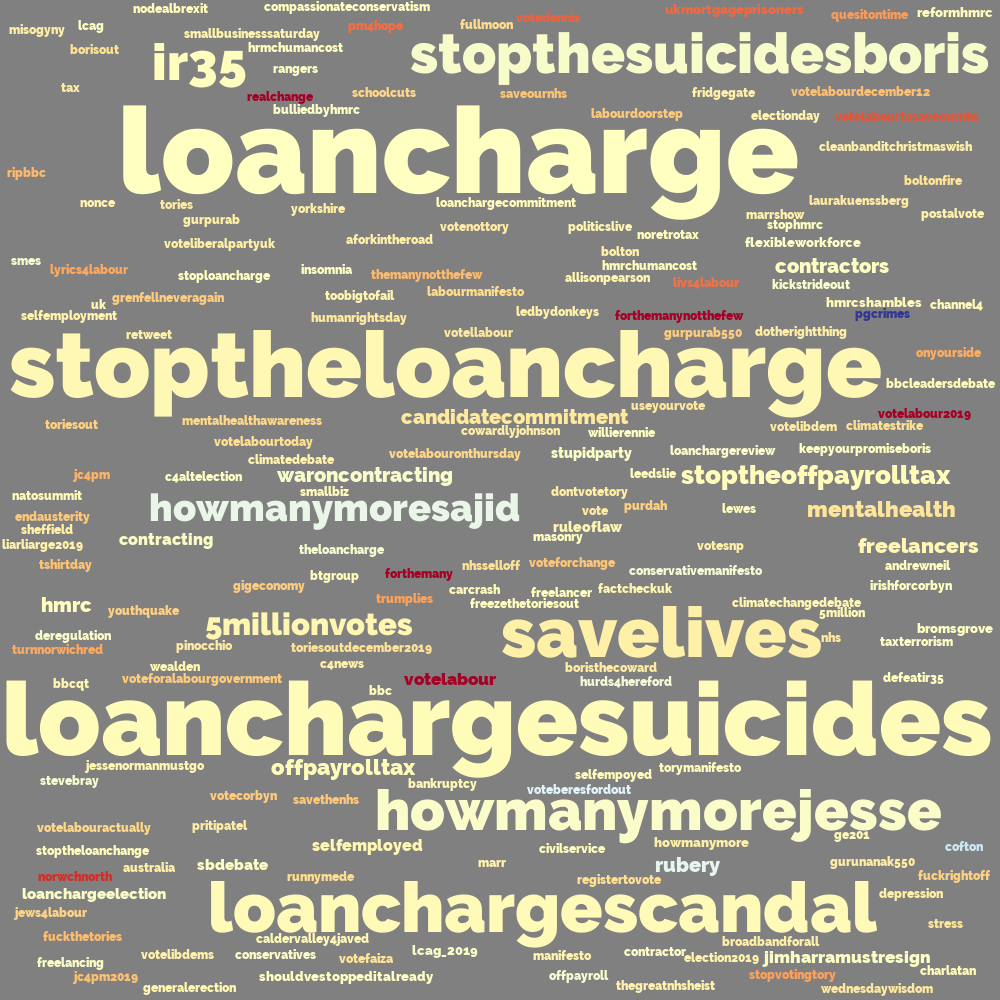}
            \caption{\tikz\draw[black, fill=lch] (0,0) circle (.75ex);~\texttt{LCH}.\label{fig:network-hashtags-LCH}}
        \end{subfigure}%
        \caption{TF-IDF weighted hashtag clouds for the different coordinated communities, representing the key issues and debates within the communities. Hashtag polarity is color-coded.}
        \label{fig:network-hashtags}
	\end{minipage}\\%
\end{figure*}

\textbf{Network interpretation.} As shown in Figure~\ref{fig:network}, the user similarity network presents a visible structure characterized by several large communities and a few smaller ones. Concerning political polarization, all users can be grouped into three main classes: Labour (red-colored), Conservative (blue-colored), and Neutral users (yellow-colored). We performed a first sanity check by comparing the structural properties of the network with political ones. In particular, colors in the network appear to be clearly separated. In other words, communities derived from network structure appear to be extremely politically homogeneous, and we do not have any cluster that contains users with markedly different colors. 
Moving forward, the Conservative cluster appears to be sharply separated from the rest of the network, while the Labour and Neutral clusters are more intertwined with one another. This interesting property of our network closely resembles the political landscape in the UK ahead of the 2019 GE. Indeed, one of the debate's main topics was Brexit, which led to a strong polarization between Conservatives and all other parties.~\cite{schumacher2019brexit}. In addition, the \textit{first-past-the-post} UK voting system also motivated anti-Tory electors to converge on the candidate of the party having the highest chances to defeat the Conservative's one in each constituency -- a strategy dubbed \textit{tactical voting}\footnote{P. Kellner, ``Tactical voting guide 2019: the 50 seats where it is vital to keep the Tories out.'' \emph{The Guardian}, 8 December 2019. Available at \url{https://www.theguardian.com/politics/2019/dec/08/tactical-voting-guide-2019-keep-tories-out-remain-voter-general-election}}. Our rich and informative network embeds and conveys these nuances.

\textbf{Coordinated communities.} Building on these promising preliminary results, we are now interested in a fine-grained analysis of the user similarity network communities. 
As mentioned before, previous works focused on threshold-based approaches and enforced restrictive edge weight filters to retain only edges with very large weights. 
Instead, in our study, the filtered user similarity network features diverse degrees of similarity and coordination, as testified by the distribution of edge weights in Figure~\ref{fig:edge-weights}. 
Hence, we are able to analyze the communities' characteristics at different levels of coordination, as opposed to cutting at a predetermined, fixed threshold, which would exclude almost the entire network from the subsequent analysis.
In detail, we carried out the analysis by applying community detection and the well-known Louvain algorithm~\cite{blondel2008fast}. 
This step in our analysis corresponds to line 1 of Algorithm~\ref{alg:comdet}. Detected communities (resolution $= 1.5$, minimum size at $t_{0}$ $= 20$) are outlined in Figure~\ref{fig:network-communities} and are briefly described in the following. Users exhibiting higher coordination with other users are assigned darker shades of color. For each community, we also computed its TF-IDF weighted hashtag cloud, as shown in Figure~\ref{fig:network-hashtags}, to highlight the debated topics.
\begin{enumerate}
    \item \tikz\draw[black, fill=con] (0,0) circle (.75ex);~\texttt{CON}: The community of Conservative users that was clearly visible in Figure~\ref{fig:network} was also detected by our community detection algorithm. It includes all major Conservative users (e.g., \hashtag{@BorisJohnson} and \hashtag{@Conservatives}), and it is characterized by a majority of hashtags supporting the Conservative Party (\hashtag{voteconservative}), its leader (\hashtag{backboris}) and Brexit (\hashtag{getbrexitdone}).
    \item \tikz\draw[black, fill=lab] (0,0) circle (.75ex);~\texttt{LAB}: Similarly, also the dense group of Labour users that we highlighted in Figure~\ref{fig:network} has been identified as a distinct community of Labours. These users are characterized by hashtags supporting the party (\hashtag{votelabour}), their leader (\hashtag{jc4pm}), and traditional Labour flags like healthcare (\hashtag{saveournhs}) and climate change (\hashtag{climatedebate}). Notably, the absence of Brexit-related keywords seems to confirm the alleged ambiguity of Jeremy Corbyn's campaign on this topic\footnote{T. Harris, ``Labour's policy of constructive ambiguity over Brexit is running out of road.'' \emph{The Telegraph}, 3 September 2019. Available at \url{https://www.telegraph.co.uk/politics/2019/09/03/labours-policy-constructive-ambiguity-brexit-running-road/}}.
    \item \tikz\draw[black, fill=tvt] (0,0) circle (.75ex);~\texttt{TVT}: The largest group of neutral users in Figure~\ref{fig:network}, tightly related to \texttt{LAB} users, was assigned to this community. These users debated topics related to liberal democrats (\hashtag{votelibdem}), anti-Tory (\hashtag{liarjohnson}), anti-Brexit (\hashtag{stopbrexit}) and to the campaigns promoting tactical voting (\hashtag{votetactically}, \hashtag{tacticalvote}).
    \item \tikz\draw[black, fill=snp] (0,0) circle (.75ex);~\texttt{SNP}: The remaining share of neutral users was assigned to this community, related to the Scottish National Party (SNP). The main hashtags used by members of this community support the party (\hashtag{votesnp}) and ask for a new referendum for the independence from the UK (\hashtag{indyref2020}). The traditional hostility of SNP against Brexit and Tories~\cite{jackson2019uk} also explains the proximity of this cluster to the \texttt{LAB} and \texttt{TVT} ones.
    \item \tikz\draw[black, fill=b60] (0,0) circle (.75ex);~\texttt{B60}: This small cluster identifies activists involved in the so-called \textit{Backto60} initiative (\hashtag{backto60}, \hashtag{50swomen}), which represents 4 million women born in the 1950s that are negatively affected by state pension age equalisation. Their instances have been addressed in the Labour manifesto, while Conservatives denied their support to the initiative despite Boris Johnson's promises\footnote{S. Smith, ``Backto60 granted leave to appeal.'' \emph{Pensions Age Magazine}, 21 January 2020. Available at \url{https://pensionsage.com/pa/Backto60-granted-leave-to-appeal.php}}. The political connections of \textit{Backto60} activists are well reflected in our network, as represented by the \texttt{B60} cluster being linked to both the \texttt{LAB} and \texttt{TVT} clusters.
    \item \tikz\draw[black, fill=ase] (0,0) circle (.75ex);~\texttt{ASE}: The tightly connected users in this cluster are all strongly leaning towards Conservatives, as also clearly visible by their connections. However, their activities are mainly devoted towards attacking the Labour party and its leader, rather than to support the Tories. As confirmed from Figure~\ref{fig:network-hashtags-ASE}, some of the most relevant hashtags of this cluster are against Labours (\hashtag{labourlies}, \hashtag{nevercorbyn}) and, in particular, are about the antisemitism allegations (\hashtag{labourantisemitism}, \hashtag{votelabourvoteracism}) that held the stage during the entire electoral campaign\footnote{G. Pogrund, J. Calavert, and G. Arbuthnott, ``Revealed: the depth of Labour anti-semitism.'' \emph{The Times}, 8 December 2019. Available at \url{https://www.thetimes.co.uk/article/revealed-the-depth-of-labour-anti-semitism-bb57h9pdz}}.
    \item \tikz\draw[black, fill=lch] (0,0) circle (.75ex);~\texttt{LCH}: Finally, the last cluster is again composed of activists, similarly to the \texttt{B60} cluster. This time activists were protesting against ``loan charge'', a tax charge introduced to contrast a form of tax avoidance based on disguised remunerations. Anti-loan charge campaigners claim that it is a retrospective taxation that, due to the abnormally long period of application, caused involved people to return unsustainable amounts, also inducing several suicides\footnote{GOV.UK, ``Disguised remuneration: guidance following the outcome of the independent loan charge review.'' \emph{GOV.UK}, 20 January 2020. Available at \url{https://www.gov.uk/government/publications/disguised-remuneration-independent-loan-charge-review/guidance}}.
\end{enumerate}
The analysis of the communities detected in our user similarity network allowed us to identify both large clusters, each corresponding to one of the major political forces involved in the election, as well as much smaller ones. The small clusters are related to highly organized activists (\texttt{B60}, \texttt{LCH}) and political campaigns (\texttt{ASE}). The previous analysis provided some first results into the presence of coordinated behaviors in the 2019 UK GE. In particular, it allowed us to uncover groups that featured at least a small degree of coordination. However, since our network embeds different degrees of coordination among its users, it still does not provide results towards the \textit{extent} of such coordination and the patterns of coordination that characterize such groups. These crucial points are tackled in the next section.

\begin{figure}[t]
    \captionsetup[subfigure]{labelformat=empty}
    \centering
    \begin{subfigure}{.24\textwidth}%
        \includegraphics[width=\textwidth]{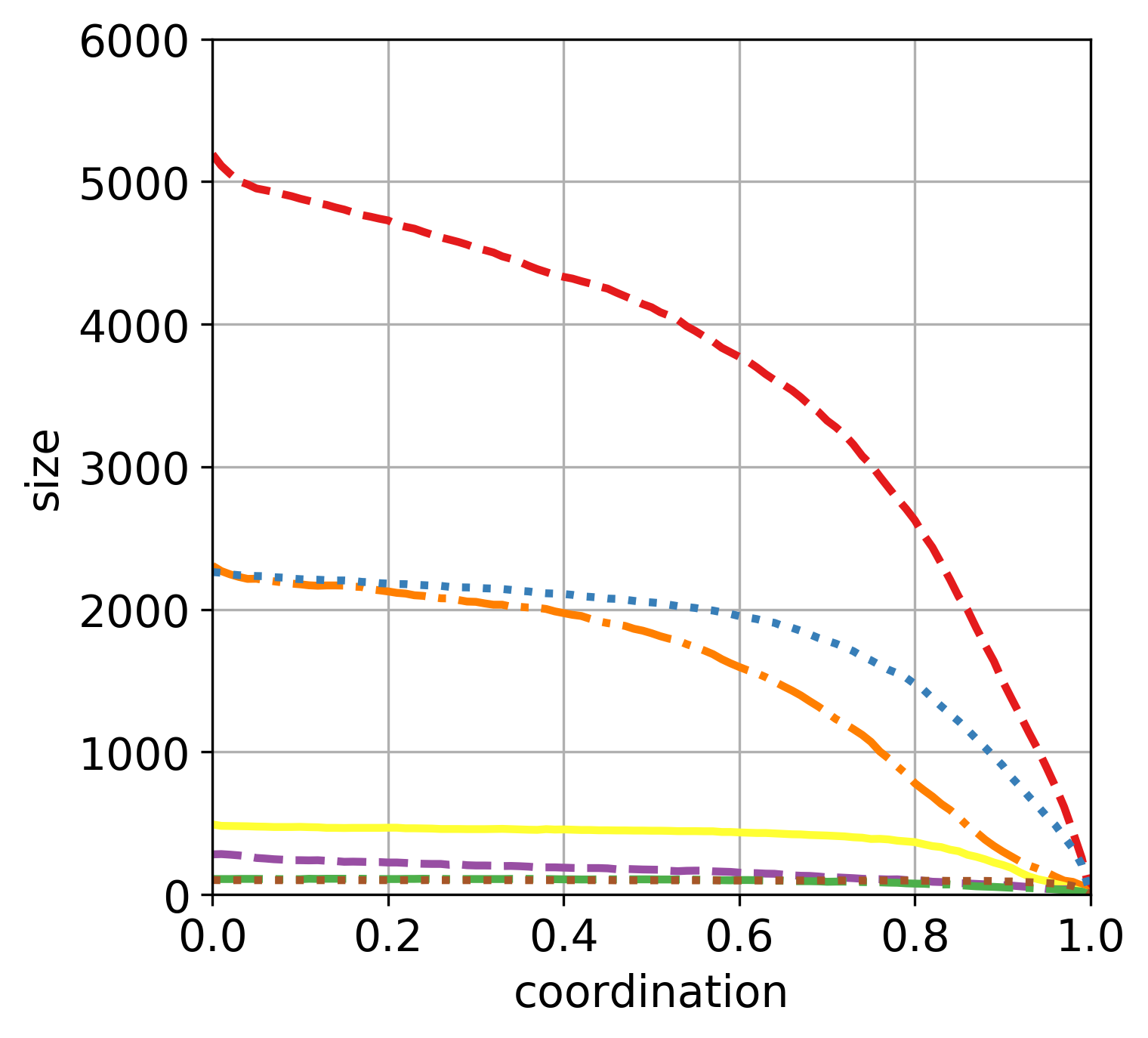}
    \end{subfigure}%
    \begin{subfigure}{.24\textwidth}%
        \includegraphics[width=\textwidth]{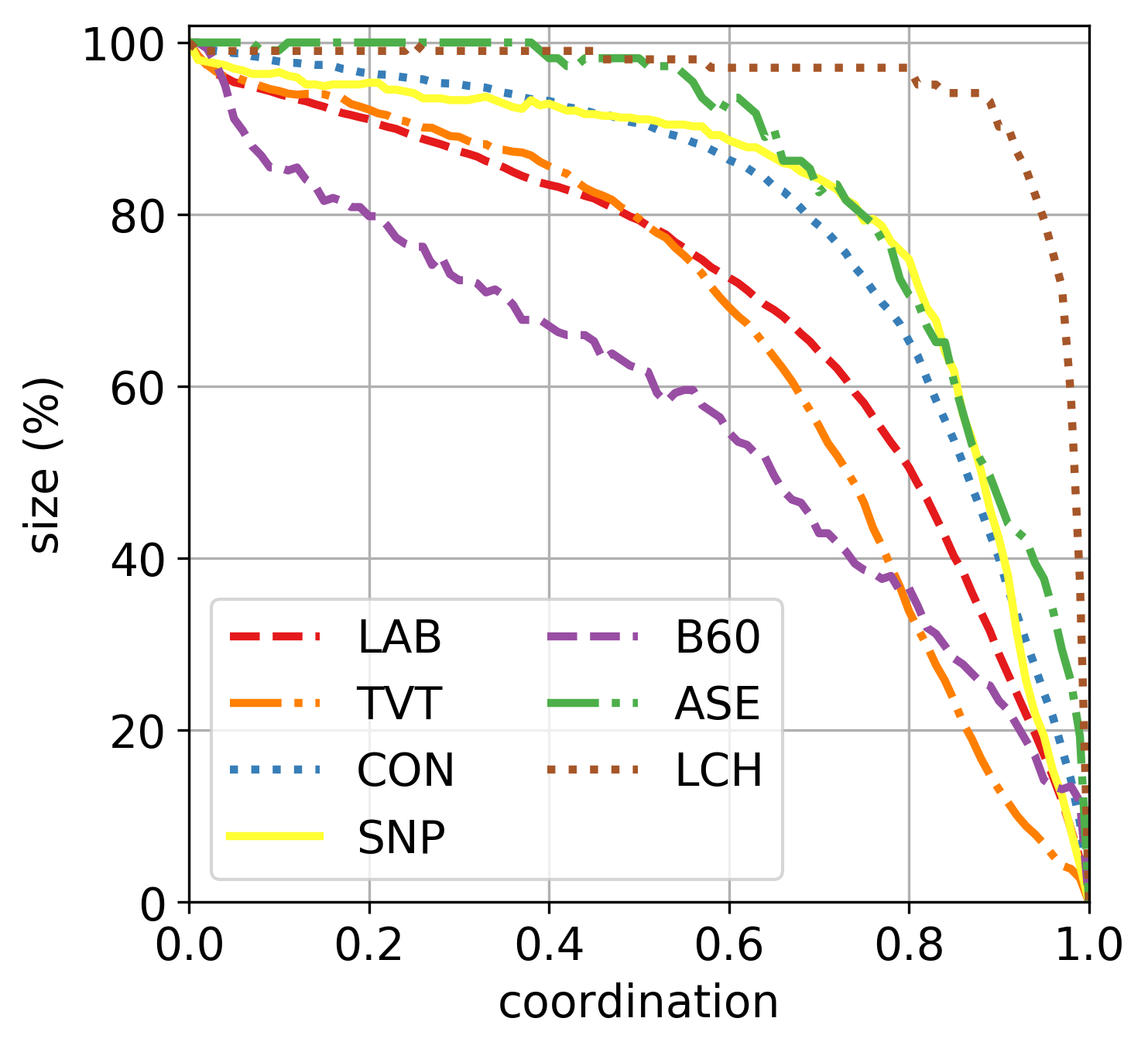}
    \end{subfigure}%
    \caption{Relationship between coordination and size of coordinated communities. Some communities are characterized by stronger coordination than others, as reflected by the plateaux that strongly-coordinated communities exhibit. This analysis enabled by the framework can provide insights for estimating the critical value that describes a given community's coordination.}
    \label{fig:coordination-vs-size}
\end{figure}

\begin{figure*}[t]
    \centering
    \begin{subfigure}{.24\textwidth}%
        \includegraphics[width=\textwidth]{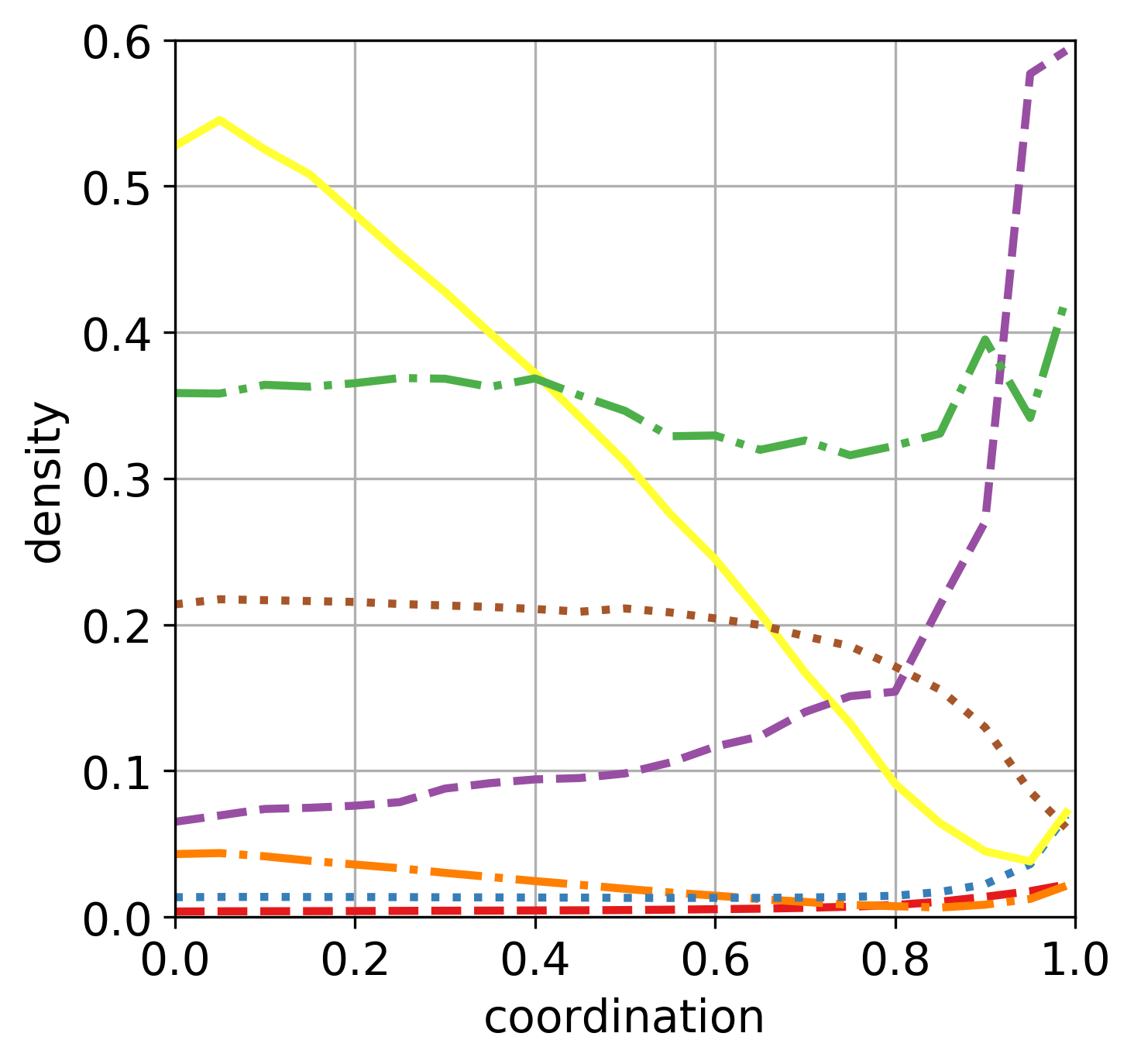}
        \caption{Density.\label{fig:coordination-vs-density}}
    \end{subfigure}\hspace{.1\textwidth}%
    \begin{subfigure}{.24\textwidth}%
        \includegraphics[width=\textwidth]{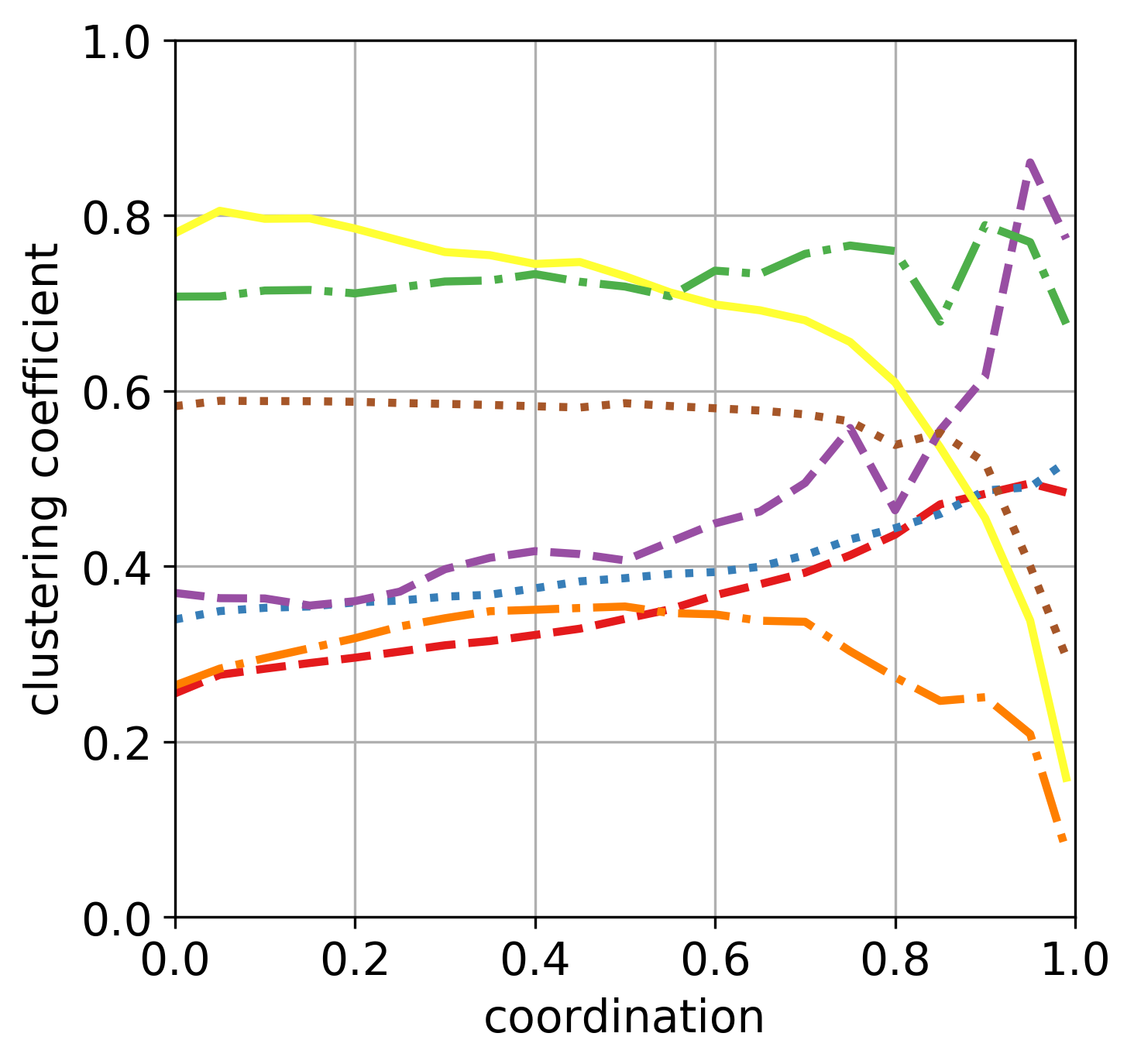}
        \caption{Clustering coefficient.\label{fig:coordination-vs-clustering}}
    \end{subfigure}\hspace{.1\textwidth}%
    \begin{subfigure}{.26\textwidth}%
        \includegraphics[width=\textwidth]{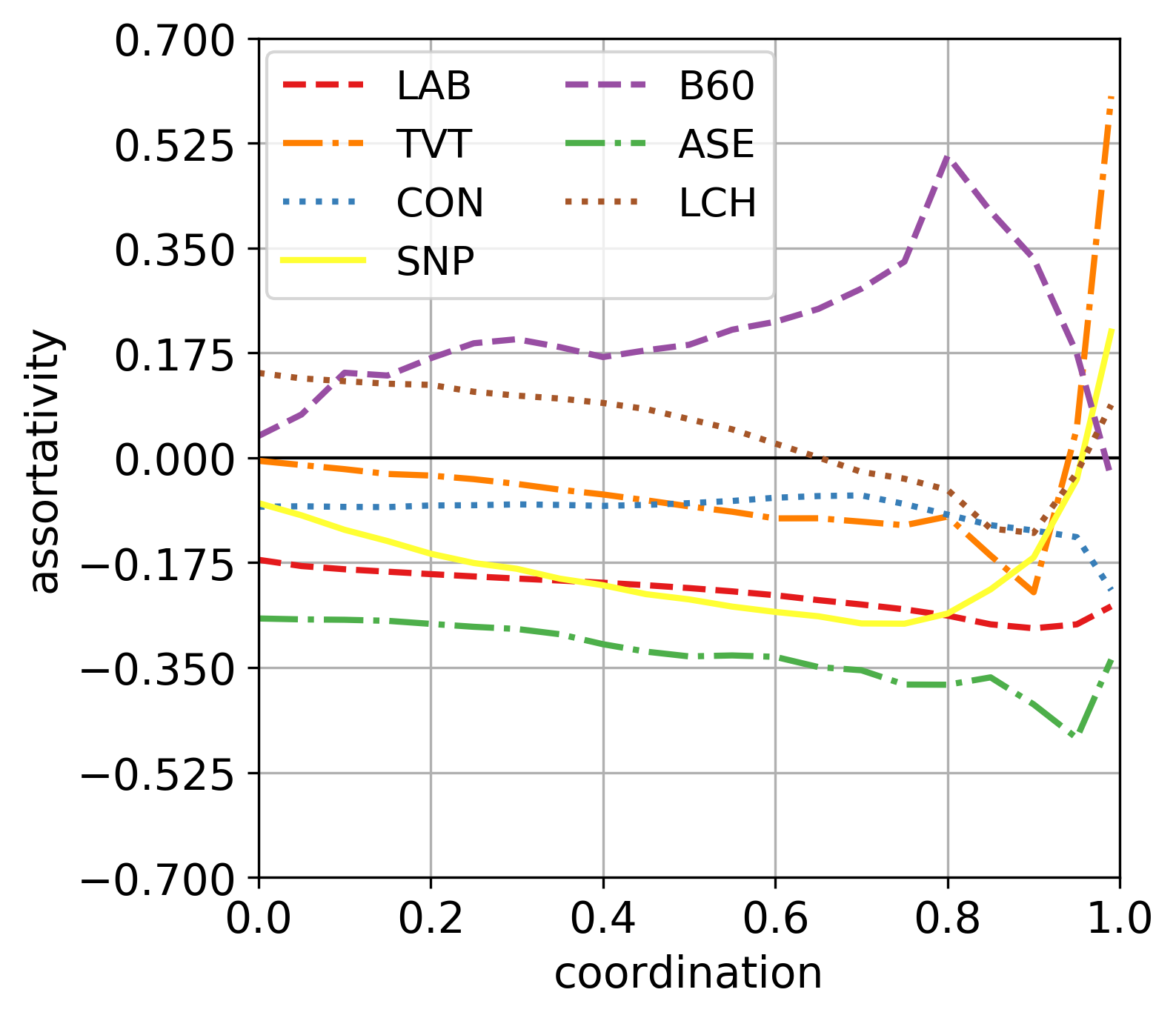}
        \caption{Assortativity.\label{fig:coordination-vs-assortativity}}
    \end{subfigure}%
    \caption{Network measures computed for each coordinated community, as a function of the extent of coordination. By studying the whole extent of coordination among users, we are able to highlight the radically different patterns of coordination that characterize different communities, as highlighted by opposite trends in given network measures.}
    \label{fig:coordination}
\end{figure*}

\section*{Analysis of coordinated behaviors}
\label{sec:analysis}
In previous work focused on threshold-based approaches, once detected, coordinated communities were visualized and manually inspected. In other words, existing pipelines for automatically studying coordinated behaviors stop at detecting coordinated communities (step~\ref{itm:step5} in our framework), without providing insights into the patterns of coordination, which are left to human analysts. Contrarily, our multifaceted analysis allows our framework to produce results for estimating the extent and investigating the patterns of coordination.

\textbf{Visual inspection.} Regarding the extent of coordination, a visual inspection of Figure~\ref{fig:network-communities} already reveals interesting insights. For instance, large communities such as \texttt{LAB} and \texttt{CON} are simultaneously characterized by a multitude of weakly-coordinated users (light-colored) and by a smaller core of strongly-coordinated ones (dark-colored). Instead, other communities only feature either weakly- or strongly-coordinated behaviors. For example, the \texttt{SNP} and \texttt{TVT} communities appear to be characterized by mildly-coordinated behaviors, with only a few strongly-coordinated users that are spread out in the network and not clustered together. On the opposite, the small communities of activists (\texttt{B60}, \texttt{LCH} and \texttt{ASE}) appear to be almost completely characterized by strongly-coordinated behaviors, as represented by small, compact, and dark-colored clusters.

\textbf{Network measures.} In the following, we formalize these intuitions and propose 
a set of network measures for quantifying them. By applying steps~\ref{itm:step5} and~\ref{itm:step6} of our framework, we 
can produce these results automatically for each uncovered coordinated community. In detail, the \textit{while}-loop in Algorithm~\ref{alg:comdet} repeatedly performs community detection on subnetworks obtained by iteratively removing edges (and the resulting disconnected nodes) based on a moving threshold on the edge weight.
Namely, we begin by removing weak edges, and we proceed with stronger ones until we have removed all edges and nodes in the network. Since edge weight is a proxy for coordination, each subnetwork we obtain with this process features a different extent of coordination, which can be measured in terms of the corresponding value of the moving threshold. In particular, given a subnetwork obtained at a specific iteration of our algorithm, we measure its coordination extent as the percentile rank\footnote{The percentile rank is the proportion of values in a distribution that a particular value is $\ge$ to.} of the corresponding moving threshold with respect to the edge weight distribution of the filtered network, shown in Figure~\ref{fig:edge-weights}. For example, a degree of coordination $ = 0.9$ means that the considered subnetwork includes only the top-10\% of strongest edges of the overall filtered network, obtained at step~\ref{itm:step4}. As a result of this approach, we can characterize the patterns of coordination in terms of standard network measures as a function of the coordination extent, for each detected community.

The first aspect we consider is the \textit{size} of coordinated communities. Figure~\ref{fig:coordination-vs-size} shows how the number and the percentage of nodes in each coordinated community changes as a function of coordination. This analysis quantifies the observations we previously derived by visual inspection. It clearly shows that some communities are characterized by stronger coordination than others. This is reflected by the plateaux that strongly-coordinated communities, such as \texttt{LCH} and \texttt{ASE}, exhibit until some large values of coordination. On the contrary, communities such as \texttt{B60}, \texttt{LAB} and \texttt{TVT} exhibit a marked decreasing trend throughout all the spectrum of coordination. This analysis is also useful for estimating a characteristic value of coordination for a given community. For instance, by using the elbow method, the \texttt{LCH} community could be described by a coordination value $\simeq 0.9$, since the vast majority of its members feature a degree of coordination $\ge$ than that. Similarly, the \texttt{ASE} community could be characterized by a coordination value $\simeq 0.55$. These results also imply that, in general, each community has its own characteristic value of coordination. Therefore, methods that apply the same arbitrary fixed threshold to all communities risk neglecting and erasing relevant patterns.

Next, we evaluate the structural properties of coordinated communities. \textit{Density} is a measure of the fraction of the actual connections between nodes in a network, with respect to all possible connections. 
This aspect is helpful towards assessing whether the most coordinated users are all linked to one another, or whether they act in different regions of their community. Results shown in Figure~\ref{fig:coordination-vs-density} highlight interesting patterns. First of all, some communities are overall more clustered than others, such as \texttt{ASE} and \texttt{LCH}. This is another indicator of strongly-coordinated behaviors. Then, we have rising and decreasing density trends. In detail, \texttt{SNP} exhibits a negative correlation between density and coordination, implying that the most coordinated users in that community are likely not colluded nor organized between themselves. On the contrary, the most coordinated members of \texttt{B60} are likely well-organized together, as shown by the density spike observed when coordination $\ge 0.8$. \textit{Clustering coefficient}, shown in Figure~\ref{fig:coordination-vs-clustering}, provides similar results with respect to density. In fact, it shows decreasing trends for \texttt{SNP}, \texttt{LCH} and \texttt{TVT}, as well as rising trends for \texttt{CON} and \texttt{LAB} and, to a much greater extent, for \texttt{B60}. Trends in density and clustering coefficient confirm that coordination $\simeq 0.9$ appears to be a representative value for \texttt{LCH}.

Finally, we evaluate the \textit{assortativity} of coordinated networks. Here, assortativity measures the extent to which nodes with a high degree are connected to other nodes with a high degree, and vice versa. Again, different patterns emerge. In particular, some coordinated communities (e.g., \texttt{ASE} and \texttt{LAB}) are moderately disassortative. This result represents a situation where a few nodes with a high degree are connected to many nodes with a low degree, realizing a network structure that is similar to a star. In turn, this highlights a coordination pattern characterized by a few hubs supported by many less important nodes -- a pattern already found to be informative when studying online manipulations~\cite{nizzoli2020charting}. Conversely, the \texttt{B60} community appears to be strongly assortative, especially when considering coordination in the region of 0.8. This finding represents a situation where many similar nodes are connected to each other, reinforcing the idea of a clique of coordinated peers. By combining all results shown in Figure~\ref{fig:coordination}, the \texttt{B60} community appears to be well-described by a coordination value $\simeq 0.8$.

\begin{figure}[t]
    \centering
    \begin{subfigure}{.24\textwidth}%
        \includegraphics[width=\textwidth]{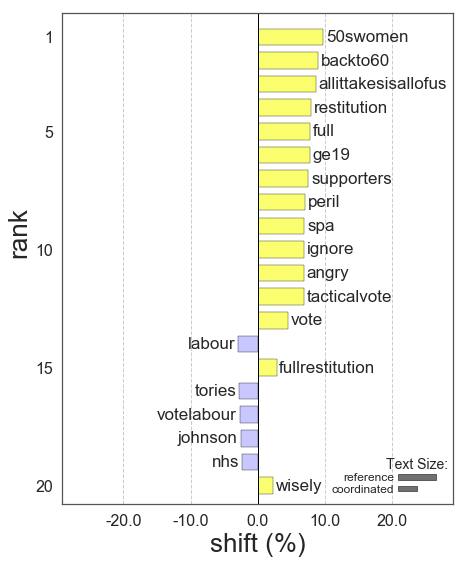}
        \caption{\tikz\draw[black, fill=b60] (0,0) circle (.75ex);~\texttt{B60}.\label{fig:narratives-B60}}
    \end{subfigure}%
    \begin{subfigure}{.24\textwidth}%
        \includegraphics[width=\textwidth]{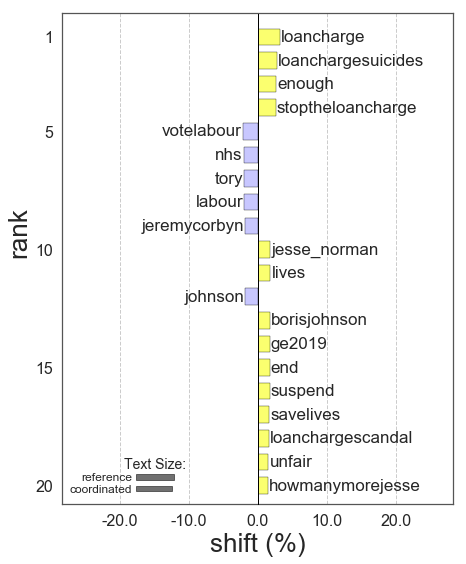}
        \caption{\tikz\draw[black, fill=lch] (0,0) circle (.75ex);~\texttt{LCH}.\label{fig:narratives-ASE}}
    \end{subfigure}%
    \caption{Application of word shift graphs for highlighting narratives that characterize coordinated communities. Words are ranked based on their contribution towards differentiating coordinated and non-coordinated users. Yellow-colored words  (right-hand  side  of  each  word  shift  graph)  are  in-formative for coordinated users, while blue-colored words(left-hand side) are informative for non-coordinated users.}
    \label{fig:narratives}
\end{figure}

\textbf{Themes and narratives.} Until now, we have only leveraged network measures to characterize coordinated communities. However, their content production can also reveal interesting insights into their preferred narratives. Here, we propose and briefly experiment with a text-based analysis that can be used to investigate the activity of coordinated groups. In particular, we are interested in highlighting the differences in the content produced by the coordinated users in a community, with respect to all other -- less coordinated -- users of that community. One way to reach our goal is by exploiting word shift graphs~\cite{gallagher2020word}, which allow comparing two corpora by highlighting those terms that mostly contribute to differentiate them. We apply word shift graphs in our context by selecting all tweets shared by members of a community as the reference corpus and all tweets shared by strongly-coordinated users in that community as the comparison corpus. Meaningful strongly-coordinated users from a community can be picked by leveraging the results of our previous network-based analyses. For instance, the \texttt{B60} community can be assigned a coordination value $\simeq 0.8$ while \texttt{LCH} can be characterized by coordination $\simeq 0.9$. Thus, in Figure~\ref{fig:narratives} we highlight content production differences between all users in \texttt{B60} and \texttt{LCH}, with respect to the users in those communities whose coordination $\ge 0.8$ and $0.9$, respectively. In figures, words are ranked based on their contribution towards differentiating coordinated and non-coordinated users. Yellow-colored words (right-hand side of each word shift graph) are informative for coordinated users, while blue-colored words (left-hand side) are informative for non-coordinated users. The informativeness of the different words towards characterizing coordinated users (i.e., their shift) is computed by means of Shannon entropy~\cite{gallagher2020word}. As shown, this analysis reveals that coordinated users embrace much more specific narratives and themes with respect to non-coordinated users. In fact, while both \texttt{B60} and \texttt{LCH} are characterized by generic Labour topics, coordinated users in those communities fight for 50s women's rights and against the loan charge tax. 

\begin{figure}[t]
    \centering
    \begin{subfigure}{.24\textwidth}%
        \includegraphics[width=\textwidth]{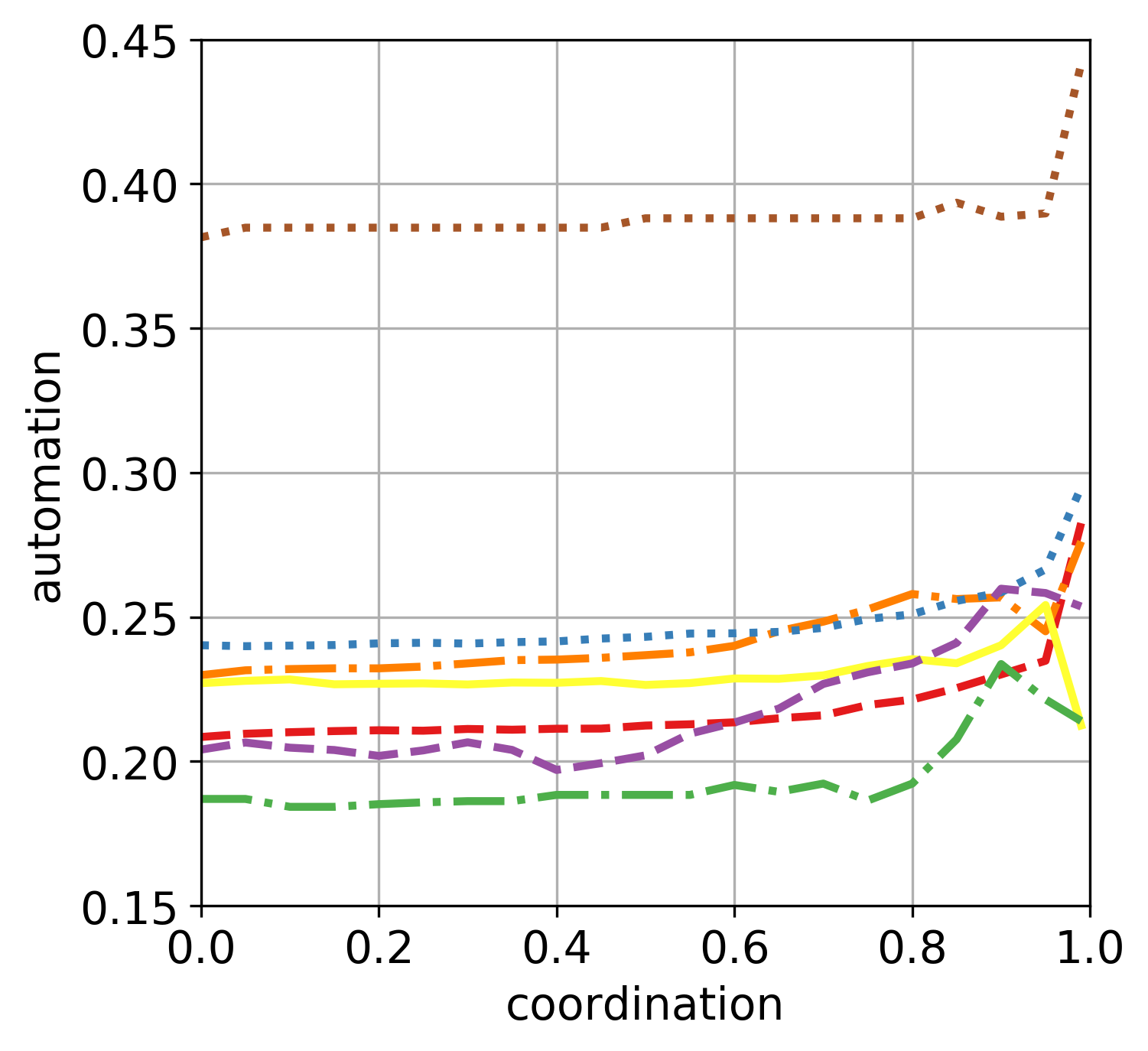}
        \caption{Mean Botometer scores.\label{fig:coordination-vs-botometer}}
    \end{subfigure}%
    \begin{subfigure}{.23\textwidth}%
        \includegraphics[width=\textwidth]{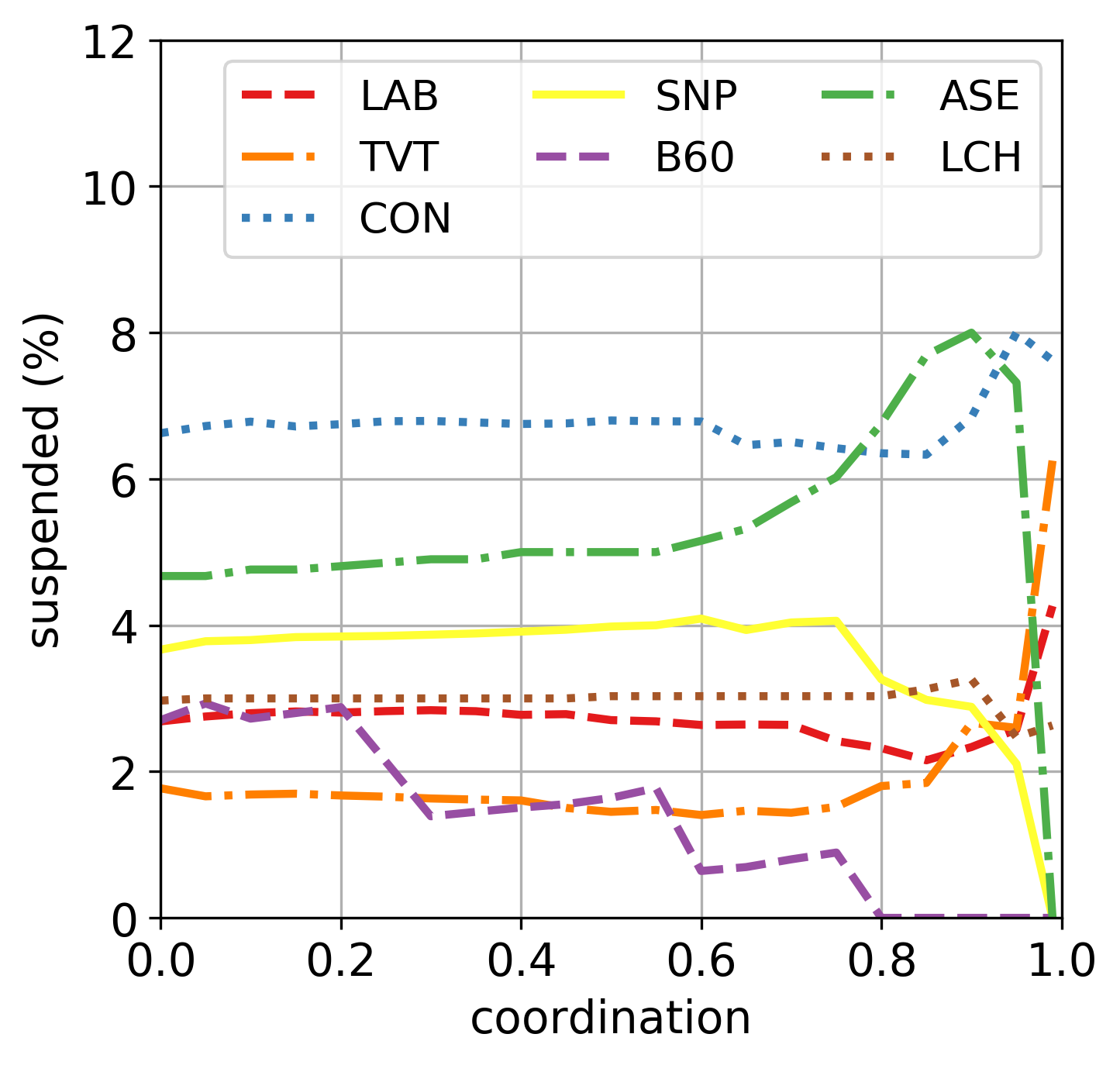}
        \caption{Suspended accounts.\label{fig:fig:coordination-vs-suspensions}}
    \end{subfigure}%
    \caption{Correlation between coordination and use of automation, in terms of bot scores estimated by Botometer, and accounts suspend by Twitter. Both indicators of possible automation appear to be two orthogonal and largely uncorrelated with respect to coordination.}
    \label{fig:coordination-vs-automation}
\end{figure}

\textbf{Use of automation.} As a last experiment on coordinated behavior, we are interested in evaluating the relationship between coordination and the use of automation. Detection of automation (e.g., social bots) has been a matter of study for years and has been one of the most widely used approaches for investigating online deception and manipulation~\cite{cresci2020decade}. Many bot detection techniques have been proposed~\cite{chavoshi2016debot,varol2017online,cresci2017tdsc,mazza2019rtbust}, but their effectiveness towards tracking IOs and CIB is still debated\footnote{Y. Roth, and N. Pickles, ``Bot or not? The facts about platform manipulation on Twitter.'' \emph{Twitter}, date. Available at \url{https://blog.twitter.com/en_us/topics/company/2020/bot-or-not.html}}. For these reasons, we compared our assessments on coordination with the automation score provided by Botometer~\cite{yang2019arming}. We used the maximum of Botometer's English and universal scores, both provided in the $[0,1]$ range, as our automation score. In addition, we also considered Twitter suspensions as an indicator of possible automation or inauthenticity. Then, similarly to our previous analyses, we reported the mean automation scores and the percentage of suspended users for the different coordinated communities as a function of coordination. Figure~\ref{fig:coordination-vs-automation} shows the results of this analysis. Automation appears to be almost completely uncorrelated with coordination. 
\begin{figure*}[ht!]
    \centering
    \begin{subfigure}{.24\textwidth}%
        \includegraphics[width=\textwidth]{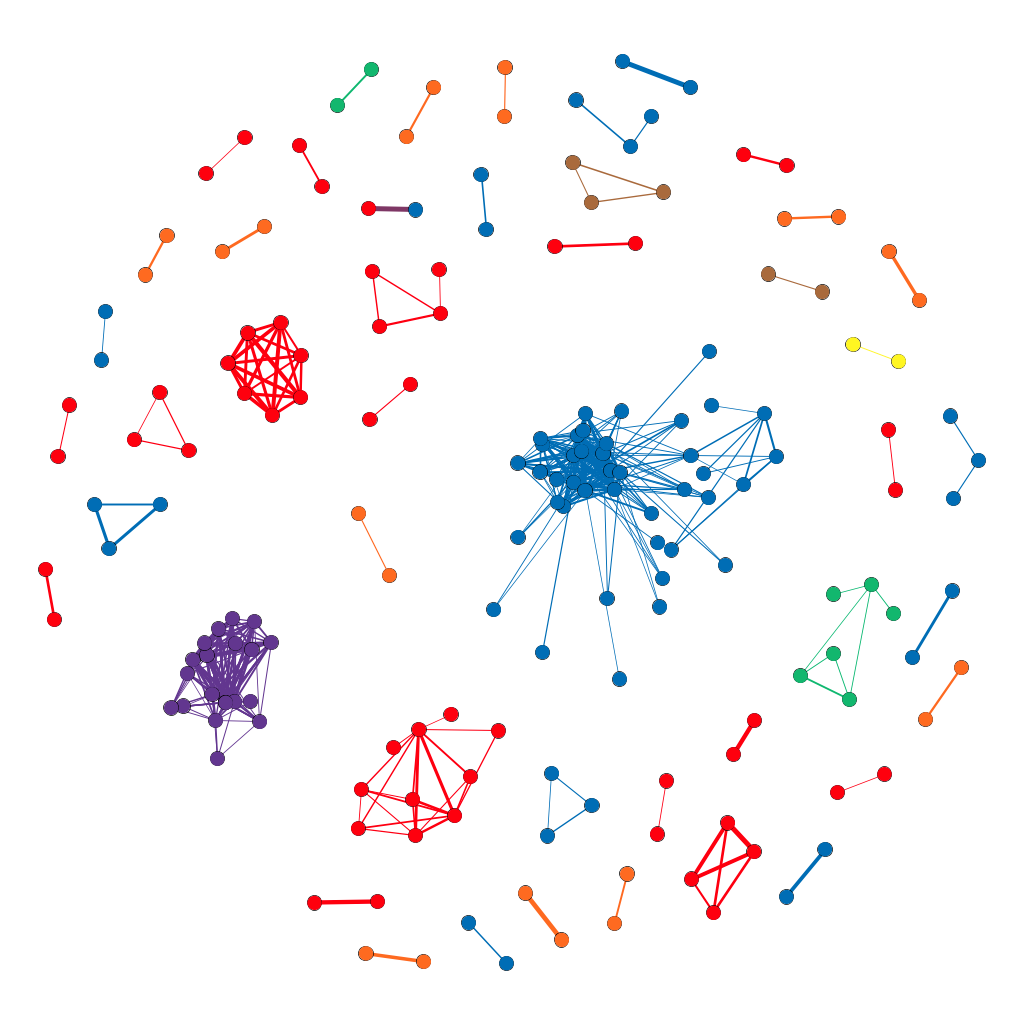}
        \caption{Edge weight threshold at 0.5.}\label{fig:comparison-pacheco-05}
    \end{subfigure}\hspace{.1\textwidth}%
    \begin{subfigure}{.24\textwidth}%
        \includegraphics[width=\textwidth]{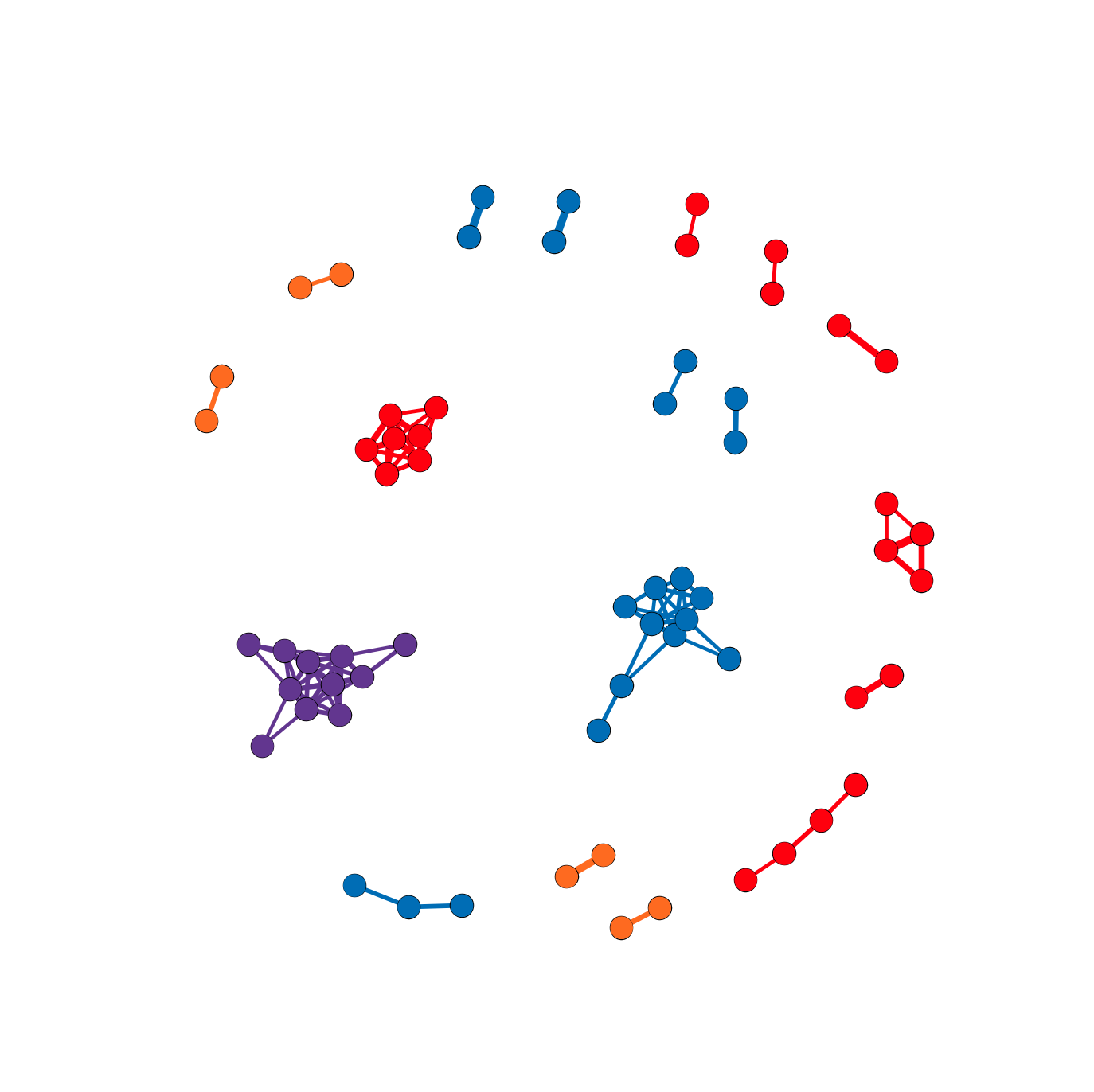}
        \caption{Edge weight threshold at 0.7.}\label{fig:comparison-pacheco-07}
    \end{subfigure}\hspace{.1\textwidth}%
    \begin{subfigure}{.26\textwidth}%
        \includegraphics[width=\textwidth]{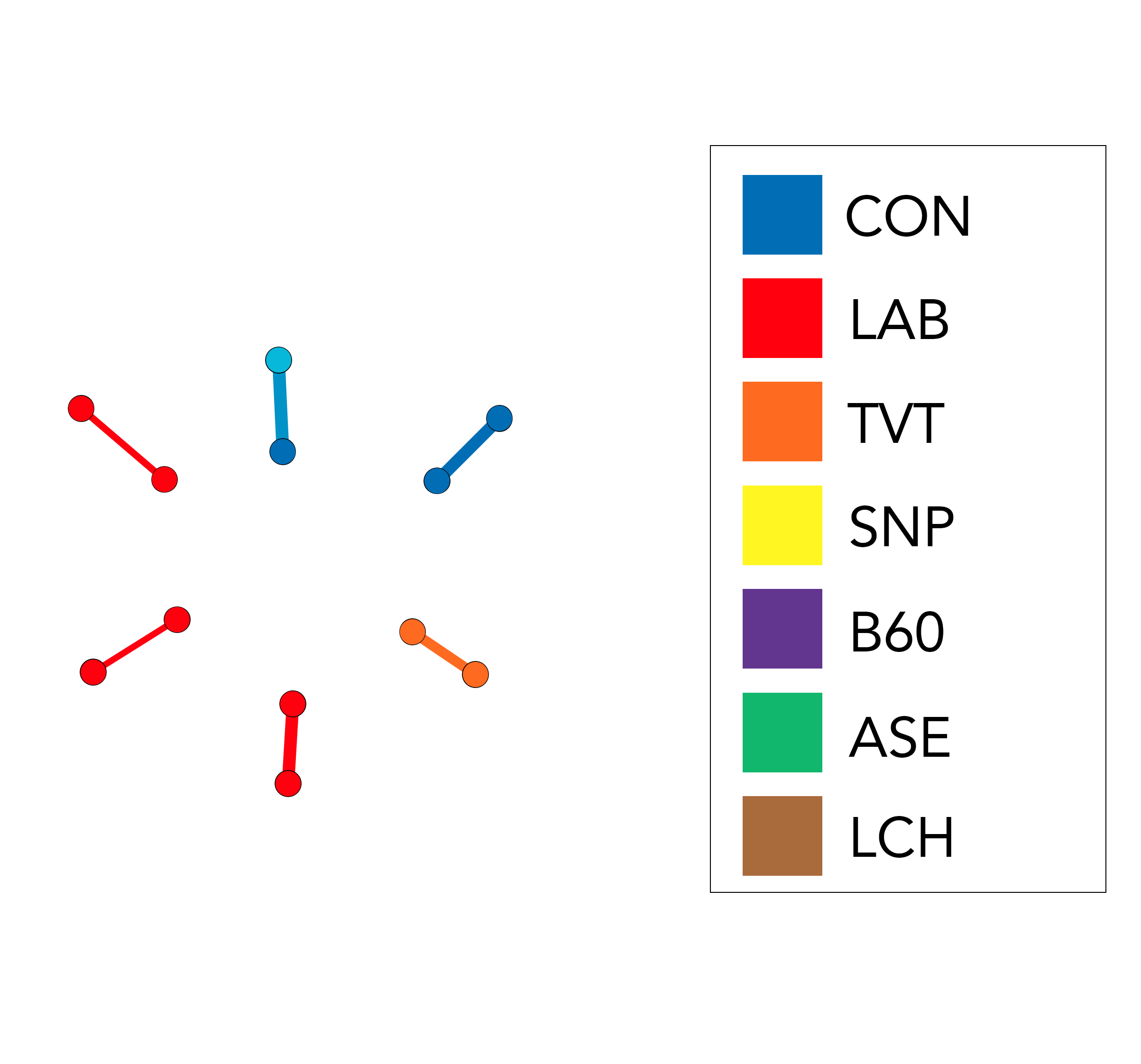}
        \caption{Edge weight threshold at 0.9.}\label{fig:comparison-pacheco-09}
    \end{subfigure}%
    \caption{Results obtained with approaches that perform simple filtering on edge weights based on fixed, arbitrary thresholds to obtain a binary distinction between coordinated and uncoordinated behaviors. As shown, these approaches capture only a small portion of the nuances and communities involved in the online debate.}
    \label{fig:comparison-pacheco}
\end{figure*}
Independently of coordination, results do not show meaningful differences between our communities, with the sole exception of \texttt{LCH} for which we measured overall higher automation scores. Other communities are more affected by Twitter suspensions, such as both clusters of Conservative users (\texttt{CON} and \texttt{ASE}).
Interestingly, we notice a marked downward trend of suspensions for the \texttt{B60} group, which might indicate an authentic, strongly-coordinated grassroots initiative.

Overall, our results confirm that coordination and automation are two different and orthogonal concepts. On the one hand, this suggests that using automation and bot detection to study CIB might be ineffective and leading to inaccurate results. On the other hand, it motivates to complement existing analyses on IOs with new results that are based on the study of coordinated behaviors.

\section*{Comparative evaluation}
\label{sec:comp}

The analysis carried out so far demonstrated how our framework embeds -- and therefore allows investigating -- the intrinsic properties of coordinated behavior. In fact, by avoiding to enforce a binary separation between coordinated \textit{vs.} uncoordinated behaviors, we were able to study this phenomenon across its entire spectrum, thus measuring the intrinsic coordination of each emerging community and characterizing it under multiple dimensions of analysis. To highlight the theoretical and practical contribution of our novel framework, in this section we compare our findings with those obtainable with previous, threshold-based approaches.

First, we consider the unfiltered user similarity network, that is what we obtained after carrying out step 3 of our framework, as outlined in Figure~\ref{fig:method}. This is where our method departs from previous approaches. Then, we set an arbitrary, very strict threshold, and we cut all those edges whose weight is below the threshold, and the resulting disconnected nodes. Finally, we visualize the obtained filtered networks. 
By comparing the results obtained in this way with those of our framework, we can comparatively evaluate the contributions of steps 4-6, which constitute the main novelty of our work.

We set the filtering threshold to $0.9$, following the value proposed in~\cite{pacheco2020uncovering} for a similar, co-retweet-based analysis. We recall that here we are directly considering edge weights, following previous approaches, whereas in our framework the coordination was measured as a percentile rank. We refer to the edge weight distribution shown in Figure~\ref{fig:edge-weights} to compare values. Figure~\ref{fig:comparison-pacheco-09} shows the resulting filtered network, where only six edges managed to satisfy the condition. On the one hand, such severe filtering ensures that all the surviving edges are undoubtedly related to coordinated behaviors. On the other, it discards a huge amount of information, whose usefulness is not limited to uncover much more accounts characterized by a certain extent of coordination. 
Conversely, by adopting a more advanced filtering technique, based on extracting the network multiscale backbone, our framework (in step 4) retained all the significant structures of the network, while keeping the amount of information manageable. Then, by surfacing coordinated groups of accounts \textit{within} the related communities through the novel coordination-aware community detection algorithm (step 5), our framework enabled us to discover possible shifts in the supported themes and narratives, and investigate the interactions with other users (step 6). Moreover, since our algorithm defines a continuous measure for the coordination extent, we were able to characterize the emerging groups with informative network measures when varying the coordination level. 
None of these results would have been possible by focusing only on the most coordinated accounts, as in previous works.

Furthermore, setting a suitable, unique threshold for the entire user similarity network, based on our intuition of what can be considered a suspiciously strong similarity, can prove problematic. In fact, if for example we measure the user behavior similarity according to the usage of the same hashtags, the resulting representation will be much less sparse than the one used in this paper, based on co-retweets. As a consequence, the same similarity threshold value that is suitable for the hashtag-based representation can be too restrictive for the co-retweet-based one. In Figures~\ref{fig:comparison-pacheco-07},~\ref{fig:comparison-pacheco-05}, we provide the filtered networks obtained for less severe values of the threshold ($0.7$ and $0.5$, respectively). As expected, lower threshold values imply larger coordinated groups. However, as confirmed by a visual comparison between the two figures, relaxing the threshold condition does not affect each coordinated group in the same way, with the \texttt{CON} cluster expanding much more than the \texttt{B60} or \texttt{LAB} ones. Conversely, our nuanced analysis allowed us to estimate the intrinsic coordination that characterizes each emerging group, according to a multifaceted set of measures.

In summary, our framework introduced many innovations compared to previous approaches, each of which contributed to obtaining more meaningful and significant results.
\section*{Conclusions}
\label{sec:conc}
We addressed the problem of uncovering coordinated behaviors in social media. We proposed a new network-based framework and applied it for studying coordinated behaviors in the 2019 UK General Election (GE). Our work has both theoretical and practical implications.

From a theoretical standpoint of fighting IOs and CIB, our framework goes beyond a binary definition of coordinated \textit{vs.} uncoordinated behaviors, allowing the investigation of the whole spectrum of coordination. We reached this goal via an improved network filtering and a coordination-aware community detection process. Our nuanced approach allowed us to uncover different patterns of coordination. We demonstrated that there is a certain extent of coordination in every online community but that not all coordinated groups are equally interesting.  Furthermore, while previous works applied simple, fixed coordination thresholds to the whole network, our approach allows estimating the degree of coordination that characterizes each different community, opening up more accurate and fine-grained downstream analyses.

From a practical standpoint, we created and shared a Twitter dataset for the 2019 UK GE. Despite smaller numbers, we found that Conservatives were overall more coordinated than Labours and that they also featured a higher degree of automation and Twitter suspensions. However, the communities with the largest degree of coordination were not supporters of the main parties but rather small groups of activists and political antagonists.

In summary, our work goes in the direction of embracing the growing complexity of critical phenomena, such as online deception and manipulation. Doing so would allow us to develop better models of our complex reality, which would give us higher chances of providing accurate and reliable results. Despite still not distinguishing inauthentic coordinated behaviors from authentic ones, our work makes a step forward in this direction by providing more nuanced and more accurate results.

\bibliographystyle{aaai}
\fontsize{9.0pt}{10.0pt}
\selectfont
\bibliography{bibliography.bib}

\end{document}